\documentclass[reqno,12pt]{article}

\usepackage{natbib}
\bibpunct{(}{)}{,}{a}{}{;}

\usepackage{xr}
\usepackage{dsfont}

\usepackage[T1]{fontenc}      
\usepackage{ae}                      
\usepackage[latin1]{inputenc}
\usepackage{lmodern}
\usepackage{amsmath}
\usepackage{amsfonts}
\usepackage[french,english]{babel}
\usepackage{amssymb}
\usepackage{dsfont}
\usepackage{array}
\usepackage{color}
\usepackage{epsfig}
\usepackage{fancybox}
\usepackage{amsthm}
\usepackage{frcursive}
\usepackage{float}
\usepackage{shorttoc}
\usepackage[colorlinks=true,citecolor=blue]{hyperref}
\usepackage{shorttoc}
\usepackage[boxruled]{algorithm2e}
\usepackage{pdfsync}
\usepackage{enumitem}
\usepackage[colorlinks=true,citecolor=blue]{hyperref}
 \usepackage{slashbox}
\usepackage{caption}
\usepackage{subcaption}
\usepackage{amsthm, amsmath, amsfonts, amssymb}%
\usepackage{appendix} 
\usepackage{chngcntr} 
\usepackage{etoolbox} 
\usepackage{lipsum} 
\usepackage{rotating}
\usepackage[normalem]{ulem}
\usepackage{pifont}
\usepackage{float}
\usepackage{multirow}

\usepackage[T1]{fontenc}      
\usepackage{ae}                      
\usepackage[latin1]{inputenc}
\usepackage{lmodern}
\usepackage{amsmath}
\usepackage{amsfonts}
\usepackage[french,english]{babel}
\usepackage{amssymb}
\usepackage{dsfont}
\usepackage{array}
\usepackage{epsfig}
\usepackage{fancybox}
\usepackage{amsthm}
\usepackage{frcursive}
\usepackage{float}
\usepackage{shorttoc}

\theoremstyle{plain} 
\newtheorem{theorem}{Theorem}[section]
\newtheorem{lemma}[theorem]{Lemma} 

\newtheorem{proposition}[theorem]{Proposition} 
\newtheorem{assumption}[theorem]{Assumption}{\bf}{\rm}%
\newtheorem{remark}[theorem]{Remark}

\newtheorem{example}[theorem]{Example}
\theoremstyle{remark}

\DeclareMathOperator{\pen}{pen}

\DeclareMathOperator{\Card}{Card}

\DeclareMathOperator{\var}{Var}

\DeclareMathOperator{\MISE}{MISE}

\DeclareMathOperator{\MISEg}{MISEg}

\DeclareMathOperator{\ISE}{ISE}

\DeclareMathOperator{\ISEstand}{ISEstand}
\DeclareMathOperator{\ISEtotal}{ISEtotal}
\DeclareMathOperator{\ISEg}{ISEg}

\newcommand{\bs}{\boldsymbol}
\newcommand{\sumin}{\displaystyle\sum_{i=1}^n}

\newcommand{\inttau}{\displaystyle\int_{0}^{\tau}}

\newcommand{\e}{{\mathrm e}}
\newcommand{\diff}{\mathrm d}

\newcommand{\expbeta}{\e^{\bs{\beta_0^TZ_i}}}

\newcommand{\s}{S(u,\bs{\beta_0})}

\newcommand{\E}{\mathbb{E}}

\newcommand{\alphainf}{||\alpha_0||_{\infty,\tau}}

\newcommand{\Lint}{\mathbb{L}^2(\mathbb{R})}

\usepackage{geometry}
\usepackage{pdfsync}

\RequirePackage{amsthm, amsmath, amsfonts, amssymb}%

\usepackage{tikz}             
\usepackage{mathrsfs}

\usepackage[boxruled]{algorithm2e}

\definecolor{darkblue}{rgb}{0.0,0.0,0.7}

\hypersetup{%
  pdfauthor = {Agathe Guilloux, Sarah Lemler and Marie-Luce Taupin},%
  pdftitle = {},%
  pdfcreator = {pdflatex},%
  pdfproducer = {pdflatex}}

\date{}

\begin{document}

\title{Adaptive kernel estimation of the baseline function in the Cox model, with high-dimensional covariates}
%
%
%
%
%


\author{Agathe Guilloux\\
\small{
Laboratoire de Statistique Th\'eorique et Appliqu\'ee,}\\
\small{Universit\'e Pierre et Marie Curie - Paris 6}\\
\small{\textit{e-mail :} \texttt{agathe.guilloux@upmc.fr}}\\
\and
Sarah Lemler\\
\small{
Laboratoire de Math\'ematiques et Mod\'elisation d'\'Evry, UMR CNRS 8071- USC INRA,}\\
\small{Universit\'e d'\'Evry Val d'Essonne, France}\\
\small{\textit{e-mail :} \texttt{sarah.lemler@genopole.cnrs.fr}}\\
\and
Marie-Luce Taupin\\
\small{Laboratoire de Math\'ematiques et Mod\'elisation d'\'Evry, UMR CNRS 8071- USC INRA,}\\
\small{Universit\'e d'\'Evry Val d'Essonne, France}\\
\small{\textit{e-mail :} \texttt{marie-luce.taupin@genopole.cnrs.fr}}\\
\small{Unit\'e MaIAGE, INRA Jouy-En-Josas, France}
}


\maketitle


\begin{abstract}

The aim of this article is to propose a novel kernel estimator of the baseline function in a general high-dimensional Cox model, for which we derive non-asymptotic rates of convergence. 
To construct our estimator, we first estimate the regression parameter in the Cox model via a Lasso procedure. We then plug this estimator into the classical kernel estimator of the baseline function, obtained by smoothing the so-called Breslow estimator of the cumulative baseline function.  We propose and study an adaptive procedure for selecting the bandwidth, in the spirit of \citet{GL2011}. We state non-asymptotic oracle inequalities for the final estimator, which reveal the reduction of the rates of convergence when the dimension of the covariates grows. \\

\textit{Keywords:} Cox's proportional hazards model; Conditional hazard rate function; Semi-parametric model; High-dimensional covariates; Counting processes; Kernel estimation; Goldenshluger and Lepski method, Non-asymptotic oracle inequalities; Survival analysis
\end{abstract}
%
%

\section{Introduction}

The Cox model, introduced by \citet{Cox}, is a regression model often considered in survival analysis to relate the distribution of a time $T$ to the values of covariates. The hazard function of $T$  is then defined by 
\begin{align}
\label{eq:CoxIntro34}
\lambda_0(t, \bs{Z}) = \alpha_0(t)\exp(\bs{\beta_0^TZ}),
\end{align}
where $\bs{Z}=(Z_1,...,Z_p)^T$ is a p-dimensional vector of covariates, $\bs{\beta_0} = (\beta_{0_1} , ..., \beta_{0_p})^T$ the vector of regression coefficients and $\alpha_0$ the baseline hazard function. 

The regression parameter $\bs{\beta_0}$ and the baseline function $\alpha_0$ are the two unknown parameters in this model. Yet, more attention has been paid to the estimation of the regression parameter than to the estimation of the baseline function.

There are good reasons for this. First, the Cox partial log-likelihood, introduced by \citet{Cox}, allows to estimate $\bs{\beta_0}$ without the knowledge of $\alpha_0$. Secondly, the regression parameter is directly related to the covariates. Therefore, in order to select the relevant covariates that explain the best the survival time, we need to estimate the regression parameter. A lot of papers deal with the problem of the estimation of $\bs{\beta_0}$, the number of covariates $p$ being large or not compared with the size of the panel $n$. When $p$ is smaller than $n$, the usual estimator of $\bs{\beta_0}$ is obtained by maximizing the Cox partial log-likelihood (see \citet{ABG} as a reference book). When the number of covariates grows, the Lasso procedure is often considered. This procedure consists in the minimization of the opposite of the $\ell_1$-penalized Cox partial log-likelihood. Asymptotic results are stated in \citet{Fan,Kong,bradic2012}. Finally, the non-asymptotic rate of convergence of the Lasso is now known to be of order  $\sqrt{\log p/n}$, see \citet{Huang2013}. 

The estimation of the baseline function $\alpha_0$ has been less studied. 
The known estimator of the baseline function is a kernel estimator, introduced by \citet{ramlau83b,ramlau83a}. We present here its form in the special case of right-censoring. Let us consider, for the moment, that we observe for $i=1,...,n$, $(X_i,\delta_i,\bs{Z_i})$, where $X_i=\min({T_{i}},{C_{i}})$, $\delta_{i}=\mathds{1}_{\{T_{i}\leq C_{i}\}}$, $T_i$ is the time of interest and $C_i$ the censoring time. The usual kernel estimator is then obtained from an estimator of the cumulative baseline function $A_0$ defined by $A_0(t)=\int_{0}^{t} \alpha_0(s)\diff s$. This estimator is called the Breslow estimator and is defined, for $t>0$, by
\begin{align}
\label{eq:BreslowCum}
\hat{A}_0(t,\bs{\hat\beta})=\sum_{i=1}^n \dfrac{\delta_i }{S_n(X_i,\bs{\hat\beta})}\ \quad \text{ with } S_n(t,\bs{\hat\beta})=\sum_{i : T_i \geq t} \exp(\bs{\hat\beta^TZ_i}),
\end{align} see \citet{ramlau83a} and \citet{ABG} for details.
From $\hat{A}_0(.,\bs{\hat\beta})$, the kernel function estimator for $\alpha_0$ is derived by smoothing the increments of the Breslow estimator. It is defined by
\begin{equation}
\label{KernelEst1}
\hat\alpha^{\bs{\hat\beta}}_h(t)=\dfrac{1}{h}\inttau K\Big(\dfrac{t-u}{h}\Big)\diff \hat A_0(u,\bs{\hat\beta}),\quad \tau\geq 0
\end{equation}
with $K:\mathbb{R}\mapsto\mathbb{R}$ a kernel with integral 1, and $h$ a positive parameter called the bandwidth. This estimator has been introduced and studied by \citet{ramlau83b,ramlau83a} within the framework of the multiplicative intensity model for counting processes, thereby extending its use to censored survival data. Consistency and asymptotic normality are proven in \citet{ramlau83a} with fixed bandwidth. 


The choice of the bandwidth in kernel estimation is  crucial, in particular when one is interested in establishing non-asymptotic adaptive inequalities. State-of-the-art methods are based on cross-validation. \citet{Ramlau1981} has suggested the cross-validation method to select the bandwidth but without any theoretical guarantees.  
For randomly censored survival data, \citet{Marron87} have shown that the cross-validation method gives the optimal bandwidth for estimating the density: the ratio between the integrated squared error for the cross-validation bandwidth and the infimum of the integrated squared error for any bandwidth almost surely converges to one. \citet{Gregoire93} has considered the cross-validated method suggested by \citet{Ramlau1981}  for the adaptive estimation of the intensity of a counting process and 
has proved some consistency and asymptotic normality results for the cross-validated kernel estimator. 
However, all the results for the adaptive kernel estimator with a cross-validated bandwidth are asymptotic.

No non-asymptotic oracle inequalities have to date been stated for the kernel estimator of the baseline function. In addition, to our knowledge, the construction of $\hat\alpha^{\bs{\hat\beta}}_h$ has not yet been considered for high-dimensional covariates. The objective of the present paper is then twofold: whatever the dimension, we aim at proposing an estimator $\hat\alpha^{\bs{\hat\beta}}$ of the baseline function, for which we can establish a non-asymptotic oracle inequality to measure its performances. The loss of prediction of $|\hat\alpha^{\bs{\hat\beta}}-\alpha_0|$ when $p$ increases will be quantified. 

To fulfill these two purposes, the idea is to estimate first the regression parameter $\bs{\beta_0}$ via a Lasso procedure applied to the Cox partial log-likelihood, then to plug this estimator in the usual kernel estimator (\ref{KernelEst1}) of the baseline hazard function and finally to select a data-driven bandwidth, following a procedure adapted from \citet{GL2011}. In the latter,  the problem of bandwidth selection in kernel density estimation is addresses and an adaptive estimator is derived, which satisfies non-asymptotic minimax bounds. This method was then considered by \citet{Doumic2012} to estimate the division rate of a size-structured population in a non-parametric setting, by \citet{BCG2013} to estimate the intensity function of a recurrent event process and by \citet{chagny2013} for the estimation of a real function via a warped kernel strategy. In the present paper, we consider it to obtain an adaptive kernel estimator of the baseline function with a data-driven bandwidth. We establish the first adaptive and non-asymptotic oracle inequality, which warrants the theoretical performances of this kernel estimator. The oracle inequality depends on the non-asymptotic control of $|\bs{\hat\beta}-\bs{\beta_0}|_1$ deduced from an estimation inequality stated by \citet{Huang2013} and extended to the case of unbounded counting processes (see \citet{GLT2015} for details).

The paper is organized as follows. In Section \ref{estimation}, we describe the two-step procedure to estimate the baseline function: first, we describe the estimation of $\bs{\beta_0}$ as a preliminary step and give the bound for $|\bs{\hat\beta}-\bs{\beta_0}|_1$ and then we focus on the kernel estimation of $\alpha_0$ and describe the adaptive estimation procedure of Goldenshluger and Lepski to select a data-driven bandwidth. In Section \ref{KernelEstimators}, we establish a non-asymptotic oracle inequalitie for the adaptive kernel estimator. The fundamental proofs are gathered in Section \ref{proof}. Lastly, a supplementary material provides some technical results needed in the proofs.

\section{Notations and preliminaries}

\subsection{Framework with counting processes}
\label{subsec:framework}

%

Consider the general setting of counting processes, which embeds the classical case of right censoring. We follow here the now classical setting of \citet{ABG} or \citet{fleming11}. For $n$ independant individuals, we observe for $i=1,...,n$ a counting process $N_{i}$, a random process $Y_{i}$ with values in $[0,1]$ and a vector of covariates $\bs{Z_i}=(Z_{i,1},...,Z_{i,p})^T\in\mathbb{R}^p$. Let $(\Omega,\mathcal{F},\mathbb{P})$ be a probability space and $(\mathcal{F}_{t})_{ t\geq 0}$ be the filtration 
defined by 
$$\mathcal{F}_{t}=\sigma\{N_{i}(s),Y_{i}(s), 0\leq s\leq t, \bs{Z_{i}}, i=1,...,n\}.$$
From the Doob-Meyer decomposition, we know that each $N_i$ admits a compensator denoted by $\Lambda_i$, such that $M_{i}=N_{i}-\Lambda_{i}$ is a $(\mathcal{F}_{t})_{t\geq 0}$  local square-integrable martingale (see \citet{ABG} for details). We assume in the following that $N_i$ satisfies an Aalen multiplicative intensity model. 
\begin{assumption}
For each $i=1,...,n$ and all $t\geq 0$, 
\begin{align}
\label{eq:aalenIntro34}
\Lambda_i(t)=\displaystyle{\int_{0}^{t}\lambda_{0}(s,\bs{Z_{i}})Y_{i}(s)\diff s},
\end{align}
where $\lambda_0(t,\bs{z})=\alpha_0(t)\e^{\bs{\beta^Tz}}$, for $\bs{z}\in\mathbb{R}^p$.
\end{assumption}

This general setting, introduced by \citet{aalen}, embeds several particular examples as censored data, marked Poisson processes and Markov processes (see \citet{ABG} for further details). This framework generalizes the case considered in \citet{ramlau83a} to unbounded counting processes and hence widens the scope of applications: we can consider the jumps of the counting to happen at times of relapse from a disease in biomedical research, times of monetization in marketing, times of blogging in social network study, etc.

\subsection{Notations}
For a real number $q\geq 1$ and a function $f:\mathbb{R}\longmapsto\mathbb{R}$ such that $|f|^q$ is integrable and bounded, we consider
\[
||f||_{\mathbb{L}^q(\mathbb{R})}=\Big(\int_\mathbb{R}|f(x)|^q\diff x\Big)^{1/q} \mbox{ and } ||f||_{\infty}=\underset{x\in\mathbb{R}}{\sup}|f(x)|.
\] 
The integrals and the supremum are restricted to the support of $f$ and for $\tau$ a positive real number, we set $||f||_{\infty,\tau}=\sup_{x\in[0,\tau]}|f(x)|$ and we simply denote by $||.||_2$ the $\mathbb{L}^2$-norm restricted to the interval $[0,\tau]$, so that 
\[
||f||_2^2=\inttau f^2(x)\diff x.
\]
For $h$ a positive real number, we define $f_h(.)=f(./h)/h$. For square-integrable functions $f$ and $g$ from $\mathbb{R}$ to $\mathbb{R}$, we denote the convolution product of $f$ and $g$ by $f*g$.
%
For a vector $\bs{b}\in\mathbb{R}^p$ and a real $q\geq 1$, we denote $|\bs{b}|_q=(\sum_{j=1}^{p}|b_j|^q)^{1/q}$.

For quantities $\gamma(n)$ and $\eta(n)$, the notation $\gamma(n)\lesssim\eta(n)$ means that there exists a positive constant $c$ such that $\gamma(n)\leq c\eta(n)$.

Finally, let $\bs{Z}\in\mathbb{R}^p$ denote the generic vector of covariates with the same distribution as the vectors of covariates $\bs{Z_i}$ of each individual $i$ and by $Z_j$ its   $j$-th component, namely the $j$-th covariates of the vector $\bs{Z}$. 


\section{Estimation procedure}
\label{estimation}
In this section, we describe the two-step procedure to estimate the baseline function. We begin by recalling the usual estimation of the regression parameter $\bs{\beta_0}$ in high-dimension. We then focus our study on the second step, which consists in the adaptive kernel estimation of the baseline function $\alpha_0$.

\subsection{Preliminary estimation of $\bs{\beta_0}$}
\label{estimationbeta}


The regression parameter $\bs{\beta_0}$ is estimated via a Lasso procedure applied to the so-called Cox partial log-likelihood introduced by \citet{Cox} and defined, for all $\bs\beta\in\mathbb{R}^p$, by
\begin{equation}
\label{CoxLikelihood}
l^*_n(\bs\beta)=\dfrac{1}{n}\sumin\inttau\log\dfrac{\e^{\bs{\beta^TZi}}}{S_n(t,\bs{\beta})}\diff N_i(t), \quad \mbox{where } S_n(\bs\beta,t)=\dfrac{1}{n}\sumin \e^{\bs{\beta^TZ_i}}Y_i(t).
\end{equation} 
The estimator $\bs{\hat\beta}$ of $\bs{\beta_0}$ is then defined by 
\begin{equation}
\label{estb0Ball3}
\bs{\hat\beta}=\underset{\bs\beta\in\mathcal{B}(0,R)}{\arg\min}\{- l^*_n(\bs{\beta})+\pen(\bs\beta)\}, \quad \mbox{with} \quad \pen(\bs{\beta})= \Gamma_n|\bs{\beta}|_1,
\end{equation}
where $\Gamma_n$ is a positive regularization parameter to be suitably chosen and $\mathcal{B}(0,R)$ is the ball defined by
\[
\mathcal{B}(0,R)=\{b\in\mathbb{R}^p : |b|_1\leq R\}, \quad \mbox{with } R>0.
\]
The ball constraint has already been considered by \citet{SVG1} or \citet{Kong}. Roughly speaking, it means that we have restrict our attention to a, possibly very large, ball around  $\bs{\beta_0}$, for which the following (very mild) assumption is needed. It is required to control the kernel estimator of the baseline function $\bs{\beta_0}$.

\begin{assumption}\
\label{ass:betaball}
We assume that $|\bs{\beta_0}|_1<+\infty$.
\end{assumption}

Concerning the covariates, we introduce the following assumption.
\begin{assumption}
\label{ass:Z}
There exists a positive constant $B$ such that for all $j\in\{1,...,p\}$, 
\[|Z_{j}|\leq B.\]
\end{assumption}
Assumption \ref{ass:Z} is a classical assumption in the Cox model to obtain oracle inequalities (see \citet{Huang2013} and \citet{bradic2012}) and seems reasonable since in practice

%
%

We know give a general version of the estimation inequality of Theorem 3.1 of \citet{Huang2013}. We refer to \citet{GLT2015} for a proof of Proposition \ref{IObeta3} in the general case.

\begin{proposition}
\label{IObeta3}
Let $k>0$, $c>0$ and $s:=\Card\{j\in\{1,...,p\}: \beta_{0_j}\neq 0\}$ be the sparsity index of $\bs{\beta_0}$. Assume that $||\alpha_0||_{\infty,\tau}<\infty$. Then, under Assumptions \ref{ass:betaball} and \ref{ass:Z}, with probability larger than $1-cn^{-k}$, we have
\begin{equation}
\label{predictionbeta}
|\bs{\hat\beta}-\bs{\beta_0}|_1\leq C(s)\sqrt{\dfrac{\log (pn^k)}{n}}
\end{equation}
where $C(s)>0$ is a constant depending on the sparsity index $s$.
\end{proposition}


In the rest of the paper, the conditions of Proposition \ref{IObeta3} will be fulfilled, so that $\bs{\hat\beta}$ satisfies Inequality (\ref{predictionbeta}). The assumption $||\alpha_0||_{\infty,\tau}<\infty$ is to found in Assumptions~\ref{ass:baseline3}.

\subsection{Estimation of $\alpha_0$}
In this subsection, we define the kernel estimator of the baseline hazard function $\alpha_0$ on which our procedure relies. We state some functional and kernel assumptions, and we describe the Goldenshluger and Lepski procedure to select a data-driven bandwidth.

\subsubsection{Kernel estimator}
We first recall the definition of the kernel estimator introduced by \citet{ramlau83a} by using kernel functions to smooth the increments of the non-parametric Breslow estimator (\ref{eq:BreslowCum}) of the cumulative intensity.

Let define $K:\mathbb{R}\rightarrow\mathbb{R}$ a kernel, namely $K$ is a function such that $\int_{\mathbb{R}}K(x)\diff x=~1$. 
The usual kernel function estimator iof $\alpha_0$ is then defined by
\begin{equation}
\label{Breslow}
\hat\alpha^{\bs{\hat\beta}}_h(t)=\dfrac{1}{nh}\sumin\inttau K\Big(\dfrac{t-u}{h}\Big)\dfrac{\mathds{1}_{\{\bar Y(u)>0\}}}{S_n(u,\bs{\hat\beta})}\diff { N_i(u)},
\end{equation} 
with 
\[
\bar Y=\dfrac{1}{n}\sumin Y_i,\quad \text{and} \quad S_n(u,\bs{\beta})=\dfrac{1}{n}\sumin\e^{\bs{\beta^TZ_i}}Y_i(u), \quad \text{for all } \bs\beta\in\mathbb{R}^p.
\]
The parameter $h>0$ is called the bandwidth. In kernel function estimation, the bandwidth has to be chosen by the user. \citet{Gregoire93} has defined a cross-validation procedure for selecting the bandwidth for the smooth estimate of intensity in the Aalen counting process. To our knowledge, all theoretical results for the kernel function estimator (\ref{Breslow}) with a bandwidth selected by cross-validation are asymptotic. The cross-validation ensures no theoretical adaptive guarantees when the size of the panel $n$ is fixed and not so large as it is the case for medical surveys where only a few patients can be observed.  
This explains our interest in providing a data-driven method to select automatically the bandwidth and obtain a kernel function estimator, for which we can warrant some non-asymptotic properties.  

In what follows, we denote the estimator under study by $\hat\alpha^{\bs{\hat\beta}}_h$ in which the Lasso estimator (\ref{estb0Ball3}) has been plugged.

\subsubsection{Functional and kernel assumptions}

Classical conditions are required on the intensity function and the kernel $K$. 
\begin{assumption}\
\label{ass:baseline3}
\begin{enumerate}[label=\textbf{(\roman*)},ref=(\roman*), leftmargin=*]
\item \label{ass:Y} For all $i\in\{1,...,n\}$, the random process $Y_i$ takes its values in $\{0,1\}$.
\item \label{ass:cS3} For $S(t,\bs{\beta_0})=\mathbb{E}[\expbeta Y_i(t)]$, there exists a positive constant $c_S$ such that,
\[
 S(t,\bs{\beta_0})\geq c_S, \quad \forall t\in[0,\tau].
\]

\item\label{ass:alpha0inf3} $||\alpha_0||_{\infty,\tau}:=\sup_{t\in[0,\tau]}\alpha_0(t)<\infty$.\\

\end{enumerate}
\end{assumption}
Assumption \ref{ass:baseline3}.\ref{ass:Y} is satisfied for all the examples quoted in the introduction. In fact, this assumption is needed to ensure that the random process $Y_i$ has a lower bound when it is nonzero. We could also have considered a modified estimator of $S_n(u,\bs{\beta})$, defined by (\ref{CoxLikelihood}), as it is usually done in the censoring case without covariates. Assumption \ref{ass:baseline3}.\ref{ass:cS3} is common in the context of estimation with censored observations (see \citet{ABG})). Assumption \ref{ass:baseline3}.\ref{ass:alpha0inf3} is required to obtain Lemma \ref{th:hatalpha-alpha0} and Theorem \ref{th:hatalphahath-alpha0} below. Nevertheless, the value $||\alpha_0||_{\infty,\tau}$ is not needed to compute the estimator (see Section \ref{sec:simus}).

The following assumptions are fulfilled by many standard kernel functions and are standard in kernel function estimation.
\begin{assumption}\
\label{ass:kernel3}
\begin{enumerate}[label=\textbf{(\roman*)},ref=(\roman*),leftmargin=*]
%

\item \label{ass:kerinf} $||K||_{\infty}=\sup_{u\in\mathbb{R}}|K(u)|<\infty$ and $||K||^2_2=\int_\mathbb{R} K^2(u)\diff u<\infty$.

\item \label{ass:nh} $nh\geq 1$ and $0<h<1$.
\item  \label{ass:ConvKernel} The kernel $K$ is of order $1$, i.e. for $j\in\{0,1,2\}$ the function $x\mapsto x^jK(x)$ is integrable and 
\[
\int_{\mathbb{R}} xK(x)\diff x=0 \quad \text{and} \quad \int_{\mathbb{R}} x^2K(x)\diff x<\infty.
\]

\end{enumerate}
\end{assumption}
Assumptions \ref{ass:kernel3}.\ref{ass:kerinf} and \ref{ass:kernel3}.\ref{ass:nh} are rather standard in kernel density estimation (see \citet{GL2011}) and has also been considered in the kernel intensity estimation by \citet{BCG2013}. Assumption \ref{ass:kernel3}.\ref{ass:ConvKernel} is only required to ensure that $K_h*\alpha_0(t)\underset{h\rightarrow 0}{\longrightarrow} \alpha_0(t)$ for all $t\in[0,\tau]$.

\begin{remark}
In this paper, we do not assume that the kernel $K$ has a compact support, by opposition to \citet{BCG2013}. The Breslow estimator (\ref{Breslow}) and the pseudo-estimator (\ref{pseudo}) are then well defined for all $t\in[0,\tau]$.
\end{remark}


%
%
%


\subsubsection{Collection of estimators}
Let $\mathcal{H}_n$ be a grid of bandwidths $h>0$, satisfying the following assumptions: 
\begin{assumption}\
\label{ass:gridHn}
\begin{enumerate}[label=\textbf{(\roman*)},ref=(\roman*),leftmargin=*]
\item\label{eq:cardHn} $\Card(\mathcal{H}_n)\leq n$.
\item \label{eq:sum1/nh} For some $a\geq 0$, $\sum_{h\in\mathcal{H}_n}\dfrac{1}{nh}\lesssim \log^a(n)$.
\item\label{eq:sumexp1/h} For all $b>0$, $\sum_{h\in\mathcal{H}_n}\exp(-b/h)<+\infty$.
\end{enumerate}
\end{assumption}
Assumptions \ref{ass:gridHn}.\ref{eq:cardHn}-\ref{eq:sumexp1/h} mean that the bandwidth collection should not be too large. 
Let us give an example of grid $\mathcal{H}_n$ that satisfies the three previous assumptions.
\begin{example}[Example of $\mathcal{H}_n$]
\label{ex:grid}
The following grid is considered in the simulations in Section \ref{sec:simus}
\[
\mathcal{H}_n=\Big\{h_j=\dfrac{1}{2^{j}},j=1,...,\lfloor\log(n)/\log(2)\rfloor\Big\},
\]
where $\varepsilon\in[0,\tau/2]$ is a small constant chosen arbitrarily as close as possible to 0.
For this grid, all the assumptions required on the bandwidths are verified. Indeed, $\Card(\mathcal{H}_n)\leq \log(n)/\log(2)\leq n$ and $\forall k=1,...,\lfloor\log(n)/\log(2)\rfloor$, we have $h_j\in[n^{-1},1]$. Moreover, Assumption \ref{ass:gridHn}.\ref{eq:sum1/nh} holds true since
\[
\sum_{j :h_j\in\mathcal{H}_n}\dfrac{1}{nh_j}=\dfrac{1}{n}\sum_{j=1}^{\lfloor\log(n)/\log(2)\rfloor}{2^{j}}=O(1).
\]
Lastly,
\[
\sum_{j :h_j\in\mathcal{H}_n}\exp(-b/h_j)=\sum_{j=1}^{\lfloor\log(n)/\log(2)\rfloor}\e^{-{b2^{j}}}=O(1)
\]
and Assumption \ref{ass:gridHn}.\ref{eq:sumexp1/h} is verified.
\end{example}

On the grid $\mathcal{H}_n$, we obtain a set of kernel estimators $\mathcal{F}(\mathcal{H}_n)=\{\hat\alpha^{\bs{\hat\beta}}_h, h\in\mathcal{H}_n\}$ of the baseline function $\alpha_0$ from the definition (\ref{Breslow}).

\subsubsection{Adaptive selection of the bandwidth}
\label{adaptive}

 We wish to automatically select a relevant bandwidth $h\in\mathcal{H}_n$, in such a way to then be able to select a kernel estimator among the set $\mathcal{F}(\mathcal{H}_n)$. As usual, we must choose a bandwidth $h$ which realizes the best compromise between the squared-bias and the variance terms. The choice should be data-driven. For this, we use a quite recent method introduced by \citet{GL2011} for  the problem of density estimation. The "Goldenshluger and Lepski method" has only been considered in two different settings: \citet{BCG2013} has applied this method to provide an adaptive kernel function estimator of the intensity function of a recurrent event process and \citet{chagny2013} has used it to estimate a real valued function from a sample of random couples (see \citet{chagny2013}). Lately, \citet{ThChagny} has also proposed a "mixed strategy", which consists in applying the "Goldenshluger and Lepski method" to select the relevant model in model selection methods for real valued function in regression models.
We consider this method to obtain an adaptive kernel function estimator of the baseline function, for which we establish a non-asymptotic oracle inequality.

Let us begin to describe the method. We can explain the idea of the method of \citet{GL2011} from an heuristic proposed by \citet{ThChagny}. We want to define $\hat\alpha^{\bs{\hat\beta}}_{\hat h^{\bs{\hat\beta}}}$ so that the risk is as close as possible as
\[
\underset{h\in\mathcal{H}_n}{\min}\{||\alpha_0-K_h*\alpha_0||^2_{2}+V(h)\}, 
\]
with 
\[
V(h)=\kappa\dfrac{\alphainf\tau}{c_S^2}\Big(\alphainf\E[\e^{2\bs{\beta_0^TZ_1}}]\tau+\E[\e^{\bs{\beta_0^TZ_1}}]\Big)\dfrac{||K||_{\Lint}^2}{nh}=O\Big(\dfrac{1}{nh}\Big),
\]
for a constant $\kappa>0$ .
In order to get closer from the bias term $||\alpha_0-K_h*\alpha_0||^2_{2}$, we replace $\alpha_0$ with an estimator $\hat\alpha^{\bs{\hat\beta}}_{h'}$ (with a fixed bandwidth $h'$), so that we obtain $||\hat\alpha^{\bs{\hat\beta}}_{h'}-K_h*\hat\alpha^{\bs{\hat\beta}}_{h'}||^2_{2}$. However, unlike the bias term, this quantity is random and thus contains some variability. We need to correct this variability by deducting the part of the variance $V(h')$. Lastly, since there are no reason to choose one bandwidth $h'\in\mathcal{H}_n$ rather than an other one, we consider the entire collection and take the maximum over this collection.

Formally, we define for $h\in\mathcal{H}_n$ 
\begin{equation}
\label{eq:Ah}
A^{\bs{\hat\beta}}(h)=\underset{h'\in\mathcal{H}_n}{\sup}\Big\{||\hat\alpha^{\bs{\hat\beta}}_{h,h'}-\hat\alpha^{\bs{\hat\beta}}_{h'}||^2_{2}-V(h')\Big\}_+
\end{equation}
where 
\begin{equation}
\label{eq:alphahhprime}
\hat\alpha^{\bs{\hat\beta}}_{h,h'}(t)=K_{h'}*\hat\alpha^{\bs{\hat\beta}}_h(t), 
\end{equation}
for any $t\geq 0$ and $h,h'$ two positive real numbers, and 
\begin{align}
\label{eq:Vh}
V(h)=\kappa\dfrac{\alphainf\tau}{c_S^2}\Big(\alphainf\E[\e^{2\bs{\beta_0^TZ_1}}]\tau+\E[\e^{\bs{\beta_0^TZ_1}}]\Big)\dfrac{||K||_{\Lint}^2}{nh},
\end{align}
for some numerical constant $\kappa>0$. A data-driven equivalent of this variance term is given in Section \ref{sec:simus}. The choice of $\kappa$ is also discussed.



From these definitions, we deduce the following choice of the bandwidth:
\begin{equation}
\label{eq:hath3}
\hat h^{\bs{\hat\beta}}=\underset{h\in\mathcal{H}_n}{\arg\min}\{A^{\bs{\hat\beta}}(h)+V(h)\}.
\end{equation}
Our adaptive kernel estimator is then $\hat\alpha^{\bs{\hat\beta}}_{\hat h^{\bs{\hat\beta}}}$. 


\section{Non-asymptotic bounds for the kernel estimator}
\label{KernelEstimators}
Now, let us state the main theorems of the chapter, which provide the first non-asymptotic oracle inequality for the adaptive kernel baseline estimator in high-dimension. 
\begin{theorem}
\label{th:hatalphahath-alpha0}
Under Assumptions \ref{ass:betaball}, \ref{ass:Z}, \ref{ass:baseline3}.\ref{ass:Y}-\ref{ass:alpha0inf3} and \ref{ass:kernel3}.\ref{ass:kerinf}-\ref{ass:ConvKernel}, if $\mathcal{H}_n$ is a finite discrete set of bandwidths such that \ref{ass:gridHn}.\ref{eq:cardHn}-\ref{eq:sumexp1/h} are satisfied, 
then there exists a constant $\kappa$ such that $\hat\alpha^{\bs{\hat\beta}}_{\hat h^{\bs{\hat\beta}}}$ defined by (\ref{eq:Vh}), (\ref{eq:Ah}) and (\ref{eq:hath3}) satisfies for $n$ large enough and $k\geq 12$:
\begin{align}
\label{eq:adaptiveOI}
\E[||\hat\alpha^{\bs{\hat\beta}}_{\hat h^{\bs{\hat\beta}}}-\alpha_0||^2_{2}]&\leq C\underset{h\in\mathcal{H}_n}{\inf}\Big\{||\alpha_h-\alpha_0||^2_{2}+V(h)\Big\}+C'(s)\dfrac{\log^{a}(n)\log(pn^k)}{n},\\
\end{align}
with 
\[
V(h)=\kappa\dfrac{\alphainf\tau}{c_S^2}\Big(\alphainf\E[\e^{2\bs{\beta_0^TZ_1}}]\tau+\E[\e^{\bs{\beta_0^TZ_1}}]\Big)\dfrac{||K||_{\Lint}^2}{nh},
\]
where $C$ is a numerical constant, $C'(s)$ a constant depending on $\tau$, $\kappa_b$, $B$, $|\bs{\beta_0}|_1$, $R$, $\alphainf$, $c_S$, $||K||_{\mathbb{L}^1(\mathbb{R})}$, $||K||_{\mathbb{L}^2(\mathbb{R})}$ and on the sparsity index $s$ of $\bs{\beta_0}$.
\end{theorem}
This inequality ensures that the adaptive kernel estimator $\hat\alpha^{\bs{\hat\beta}}_{\hat h^{\bs{\hat\beta}}}$ automatically makes the squared-bias/variance compromise. The selected bandwidth $\hat h^{\bs{\hat\beta}}$ is performing as well as the unknown oracle:
\[
h^{\bs{\hat\beta}}_{oracle}:=\underset{h\in\mathcal{H}_n}{\arg\min}\E[||\hat\alpha^{\bs{\hat\beta}}_h-\alpha_0||^2_{2}],
\]
up to the multiplicative constant $C$ and up to a remaining term of order $\log^{a}(n)\log(pn^k)/n$, which is negligible. 
%
In Inequality (\ref{eq:adaptiveOI}), the infimum term is classic in such oracle inequalities for kernel estimators: the bias term $||\alpha_h-\alpha_0||^2_{2}$ decreases when $h$ decreases and the variance term $V(h)$ increases when $h$ decreases. The remaining terms are of order 
\[
\dfrac{\log^{a}(n)\log(pn^k)}{n}=\dfrac{k\log^{a+1}(n)}{n}+\dfrac{\log^a(n)\log(p)}{n}. 
\]
\citet{chagny2013}, in the context of an additive regression model, has established an oracle inequality for the kernel estimator of the real-value regression function with a remaining term of order $1/n$. In the context of the estimation of the intensity of a recurrent event process observed under a standard censoring scheme but without covariates, \citet{BCG2013} have a logarithm term which appears in their oracle inequality with a remaining term of order $\log^{1+a}(n)/n$ instead of the expected $1/n$. This logarithm term comes from the control in $\log(n)/n$ between the distribution function $G$ and its modified Kaplan-Meier estimator $\hat G$, which appears in the kernel intensity estimator. The exponent $a$ in the remaining term arises from Assumption \ref{ass:gridHn}.\ref{eq:sum1/nh}, which is needed for the control of the difference between the kernel intensity estimator involving $\hat G$ and a pseudo-estimator that does not depend of $\hat G$. As well as \citet{BCG2013}, our kernel estimator depends on an other estimator, so that we need Assumption \ref{ass:gridHn}.\ref{eq:sum1/nh} in order to control the difference between the kernel estimator (\ref{Breslow}) and the pseudo-estimator (\ref{pseudo}). If our kernel estimator had not involved another estimator, we would have considered condition $\sum1/h\leq k_0n^a_0$, as in \citet{chagny2013}, instead of Assumption \ref{ass:gridHn}.\ref{eq:sum1/nh}. The term in $\log(p)/n$ in the remaining term comes from the control of $|\bs{\hat\beta}-\bs{\beta_0}|_1$ given by Proposition \ref{IObeta3}. This term is typical of the estimation of the regression parameter $\bs{\beta_0}$ when the number of covariates is large. There is no hope of capturing up to usual rates in this high-dimensional setting, but the loss in the variance term is only of order $\log p/n$.

When we assume that the counting processes $N_i$ are bounded for $i=1,...,n$, the variance term $V(h)$ is simpler and has the same form as the variance term in \citet{BCG2013}. In this particular case, Theorem \ref{th:hatalphahath-alpha0} takes the following form.
\begin{theorem}
\label{OI:bounded}
Under the same assumptions as in Theorem \ref{th:hatalphahath-alpha0} and assuming also that there exists $c_\tau>0$, such that $N_i(t)\leq c_\tau$ almost surely for every $t\in[0,\tau]$ and $i\in\{1,...,n\}$,
there exists a constant $\kappa$ such that $\hat\alpha^{\bs{\hat\beta}}_{\hat h^{\bs{\hat\beta}}}$ defined by (\ref{eq:Ah}), (\ref{eq:hath3}) and 
\begin{equation}
\label{eq:Vh_bounded}
V_b(h)=\kappa \dfrac{c_\tau\tau\alphainf}{c_S}\dfrac{||K||^2_{\mathbb{L}^2(\mathbb{R})}}{nh},
\end{equation}
satisfies for $n$ large enough:
\begin{align}
\label{eq:adaptiveOI}
\E[||\hat\alpha^{\bs{\hat\beta}}_{\hat h^{\bs{\hat\beta}}}-\alpha_0||^2_{2}]\leq \tilde{C}\underset{h\in\mathcal{H}_n}{\inf}\Big\{||\alpha_h-\alpha_0||^2_{2}+V_b(h)\Big\}+\tilde{C'}(s)\dfrac{\log^{a}(n)\log(np)}{n}\\
\end{align}
where $\tilde{C}$ is a numerical constant, $\tilde{C'}(s)$ a constant depending on $\tau$, $c_\tau$, $B$, $|\bs{\beta_0}|_1$, $R$, $\alphainf$, $c_S$, $||K||_{\mathbb{L}^1(\mathbb{R})}$, $||K||_{\mathbb{L}^2(\mathbb{R})}$ and on the sparsity index $s$ of $\bs{\beta_0}$.

\end{theorem}
The proof of Theorem \ref{OI:bounded} is close to the one of Theorem \ref{th:hatalphahath-alpha0} and we refer to \citet{LemlerThese} for the details.

\section{Applications}
\label{sec:simus}

\subsection{Simulation study}

The aim of this subsection is to illustrate the behavior of the kernel estimator $\hat\alpha^{\bs{\hat\beta}}_{\hat h^{\bs{\hat\beta}}}$ of the baseline function in the case of right censoring and to compare it with the usual kernel estimator with a bandwidth selected by cross-validation introduced by \citet{ramlau83a}. 

\subsubsection{Simulated datas: censored data.}

We consider a cohort of size $n$ and $p$ covariates simulated according to the Cox model (\ref{eq:CoxIntro34}) with  right censoring and
with regression parameter $\bs{\beta_0}$  chosen as a vector of dimension $p$, defined by 
\[
\bs{\beta_0}=(0.1,0.3,0.5,0,...,0)^T\in\mathbb{R}^p,
\]
for various $p\geq3$. Several choices of $n$ and $p$ have been considered, with sample size $n$ taking  values $n=200$ and $n=500$ and $p$ varying between $p=\sqrt{n}$, being $15$ and $22$ respectively and $p=n$, referred to as the high-dimension case. 
For each $n$ and $p$, the design matrix $\bs{Z}=(Z_{i,j})_{1\leq i\leq n, 1\leq j\leq p}$ is simulated independently from a uniform distribution on $[-1,1]$ and survival times $T_i$, $i=1,...,n$ are simulated according to Weibull distributions $\mathcal{W}(1.5,1)$ and $\mathcal{W}(0.5,2)$. Hence, the associated baseline function has  the form $\alpha_0(t)=a\lambda^at^{a-1}$, where $a$ and $\lambda$ stand for parameters in $\mathcal{W}(a,\lambda)$. 
The censoring times $C_i$, for $i=1,...,n$,  are simulated independently from the survival times via an exponential distribution $\mathcal{E}(1/\gamma\E[T_1])$, where $\gamma$ is adjusted to the chosen rate of censorship: $\gamma=4.5$ for 20\% of censorship and $\gamma=1.2$ for 50\% of censorship.

The time $\tau$ of the end of the study is taken as the quantile at $90\%$ of $(T_i\wedge C_i)_{i=1,...,n}$. For $i=1,...,n$, we compute the observed times $X_i=\min(T_i,\tilde C_i)$, where $\tilde C_i=C_i\wedge\tau$ and the censoring indicators $\delta_i=\mathds{1}_{T_i\leq C_i}$. The definition of $\tilde C_i$ ensures that there exist some $i\in\{1,...,n\}$ for which $X_i\geq \tau$, so that all estimators are defined on the interval $[0,\tau]$ and it prevents from a certain edge effect. Each sample $(\bs{Z_i}, T_i, C_i, X_i, \delta_i, i=1,...,n)$ is repeatedly simulated $N_e=100$ times.

The compared estimators of the baseline hazard function are both constructed with the Epanechnikov kernel, defined by
\begin{equation*}
K(u)=\dfrac{3}{4}(1-u^2)\mathds{1}_{\{|u|\leq 1\}}.
\end{equation*}
In both cases we plugg the  Lasso regression parameter estimator  $\bs{\hat\beta}$ defined by (\ref{estb0Ball3}) and implemented from the R-package \textit{glmnet}.

We compare two procedures for the data-driven choice of $h$:  the Goldenshluger and Lepski method with the selected bandwidth denoted by $\hat{h}^{\bs{\hat\beta}}_{GL}$
and the cross-validation with the selected bandwidth denoted by $\hat{h}^{\bs{\hat\beta}}_{CV}$.

\subsubsection{The Goldenshluger and Lepski method}



The adaptive bandwidth selection method, we consider here, is based on the grid of bandwidths $\mathcal{H}_n$ defined in Example \ref{ex:grid} by
\[
\mathcal{H}_n=\{1/2^k, k=0,...,\lfloor \log(n)/\log(2)\rfloor\}.
\]
In our procedure (\ref{eq:Ah}), the variance term $V(h)$ involves unknown quantities, so we consider a data-driven equivalent of it and use that the right-censoring context implies that the counting processes are bounded. Hence we implement the following procedure:
\[
\hat{h}^{\bs{\hat\beta}}_{GL}= \underset{h\in\mathbb{H}_n}{\arg\min}\{A^{\bs{\hat\beta}}(h)+\hat{V}^{\bs{\hat\beta}}_b(h)\},
\]
where, for any $t\geq 0$ and $h,h'$ two positive real numbers, 
\[
A^{\bs{\hat\beta}}(h)=\underset{h'\in\mathcal{H}_n}{\sup}\Big\{||\hat\alpha^{\bs{\hat\beta}}_{h,h'}-\hat\alpha^{\bs{\hat\beta}}_{h'}||^2_{2}-\hat{V}^{\bs{\hat\beta}}_b(h')\Big\}_+, \quad\quad \hat\alpha^{\bs{\hat\beta}}_{h,h'}(t)=K_{h'}*\hat\alpha^{\bs{\hat\beta}}_h(t), 
\]
and 
\[
\hat V^{\bs{\hat\beta}}_b(h)=\kappa'\dfrac{||\hat\alpha^{\bs{\hat\beta}}_{\max(h)}||_{\infty,\tau}||K||^2_{\mathbb{L}^2(\mathbb{R})}}{nh}.
\]
In the variance term $\hat V^{\bs{\hat\beta}}_b(h)$,  we have replaced the true unknown function $\alpha_0$ by an estimator  $\hat\alpha^{\bs{\hat\beta}}_{\max(h)}$ computed for the largest bandwidth $h$ in the grid $\mathcal{H}_n$ (see \citet{BCG2013}). The numerical constant $\kappa'$ is a universal constant that we tuned from the comparison of the MISEs for several candidate values in the range $10^{-4}-1000$, and for the two different distributions $\mathcal{W}(1.5,1)$ and $\mathcal{W}(0.5,2)$. We take $\kappa'=1$.



\subsubsection{Cross-validation method}
The bandwidth $\hat{h}^{\bs{\hat\beta}}_{CV}$ selected by cross-validation is defined by:
\[
\hat h^{\bs{\hat\beta}}_{CV}=\underset{h}{\arg\min}\Bigg\{\E\inttau(\hat\alpha^{\bs{\hat\beta}}_h(t))^2\diff t-2\sum_{i\neq j}\dfrac{1}{h}K\Big(\dfrac{X_i-X_j}{h}\Big)\dfrac{\delta_i}{\bar Y(X_i)}\dfrac{\delta_j}{\bar Y{(X_j)}}\Bigg\},
\] 
where $\bar Y=\sum_{i=1}^n\mathds{1}_{\{X_i\geq t\}}$.


%

\subsubsection{Performances}
The performances of these two estimators are evaluated via
different Integrated Squared Errors ($\ISE$). For some function $\alpha\in\mathbb{L}^2([0,\tau])$ the standard $\ISE$ and the total $\ISE$ are respectively defined by
\begin{align}
\nonumber
\ISEstand(\alpha)&=\inttau(\alpha(t)-\alpha_0(t))^2\diff t,\\
\nonumber
\ISEtotal(\alpha,\bs\beta)&=\dfrac{1}{n}\sumin\inttau(\alpha(t)\e^{\bs{\beta^TZ_i}}-\alpha_0(t)\e^{\bs{\beta_0^TZ_i}})^2\diff t.
\end{align}
The associated Mean Integrated Squared Errors are defined by $\MISEg(\alpha)=\E[\ISEg(\alpha)]$, for g$=$stand  or total, where the expectation is taken on $(T_i, C_i, \bs{Z_i})$ (for sake of simplicity, we write $\MISEg(\alpha)$ even if the $\MISE$ depends on $\bs\beta$). We obtain an estimation of the different $\MISE$ by taking the empirical mean for the $N_e=100$ replications.

\subsubsection{Results}

Table \ref{tab:MISEempCensure} gives the two empirical $\MISE$s of the kernel estimators with a bandwidth selected either by cross-validation or by the Goldenshluger and Lepski method for a Lasso estimator of the regression parameter and survival times that are distributed from $\mathcal{W}(1.5,1)$, in different censoring situation. We consider the results for two rates of censoring: a usual rate of 20\% of censoring and large rate of 50\% of censoring.

%
%

\begin{table}
\centering

\begin{tabular}{|c|c|c|c||c|c||c|c||c|c|c|c}
\hline
\multicolumn{2}{|c|}{ {\backslashbox{Dimensions}{MISEs}}}   & \multicolumn{4}{|c||}{20\%}  & \multicolumn{4}{|c|}{50\%}      \\  
    \cline{3-10}
  \multicolumn{2}{|c|}{}  &   \multicolumn{2}{|c||}{MISEstand} &  \multicolumn{2}{|c||}{MISEtot}  & \multicolumn{2}{|c||}{MISEstand}   &  \multicolumn{2}{|c|}{MISEtot }  \\ 
\hline   
  \multirow{2}{*}{$n=200$} &  $p=15$ & 0.014  & 0.017    & 0.080           &    0.082       &   0.023 &   0.029               &  0.104 &    0.120    \\
   \cline{2-10}                                        
                                          &   $p=500$  &   0.013  & 0.016        &   0.117 & 0.117  &  0.022&  0.026      &0.152&  0.154   \\ 
                                        
\hline  
\multirow{2}{*}{$n=500$} &  $p=22$ &  0.009 & 0.007            & 0.038 &  0.035       &   0.011 &0.012          & 0.055 & 0.056   \\
                                            \cline{2-10}
                                          &   $p=1000$  & 0.008 & 0.008  &  0.068  & 0.064  & 0.011 & 0.013     & 0.094  & 0.096       \\

\hline
\end{tabular}
\caption{$\MISE$s of the kernel estimators with a bandwidth selected by the Goldenshluger and Lepski method (first column for each $\MISE$) and with a bandwidth selected by cross-validation of the baseline function with a Lasso estimator of the regression parameter, given two rates of censoring: 20\% and 50\% of censoring.}
%
%
%
%
%
%
%
\label{tab:MISEempCensure}
\end{table}

As expected, witht both procedures, the $\MISE$s are degraded when the censoring rate increases. 
When we compare the standard and total
 $\MISE$s, the results are rather good for the standard $\MISE$. 
%
This is  consistent, since the total $\MISE$ measures the performances of the complete intensity estimators $\hat\lambda(t,\bs{Z})=\hat\alpha^{\bs{\hat\beta}}(t)\e^{\bs{\hat\beta^TZ}}$, including the error coming from $\bs{\hat{\beta}}$, whereas the standard $\MISE$s measures the performances of the estimators of the baseline function.  Therefor

One can see that $\MISE$s resulting from the two procedure are very similar, with very rather good  results with our procedure.

In Table \ref{tab:MISEempLois}, we give the standard $\MISE$ of the kernel estimators with a bandwidth selected either by cross-validation or by the Goldenshluger and Lepski method for different distributions of the survival times. We observe that the kernel estimator with a bandwidth selected by the Goldenshluger and Lepski method performs better than the one with a bandwidth selected by cross-validation for the two Weibull distributions.

\begin{table}[htp!]
\centering
\begin{tabular}{|c|c|c|c||c|c|}
\hline
\multicolumn{2}{|c|}{\backslashbox{Dimensions}{Distributions}}  &    \multicolumn{2}{|c||}{$\mathcal{W}(1.5,1)$ } & \multicolumn{2}{|c|}{$\mathcal{W}(0.5,2)$}    \\
\hline
  \multirow{2}{*}{$n=200$} &$p=15$  & 0.056  & 0.088  &1.02    &  1.561     \\
  \cline{2-6}
                                          & $p=200$ &0.06   &0.085    &    0.923    & 1.556              \\
 \hline
\multirow{2}{*}{$n=500$}  &$p=22$    &0.025    &  0.037  &1.006& 1.521  \\
\cline{2-6}
                                        & $p=500$ &0.027     &  0.033 &  1.098 &  1.515       \\                                        
\hline
\end{tabular}

%
\caption{MISEs for the kernel estimators with a bandwidth selected by the Goldenshluger and Lepski method (first column for each distribution) and with a bandwidth selected by cross-validation (second column for each distribution),
with a Lasso estimator of the regression parameter for two different Weibull distributions of the survival times.}
\label{tab:MISEempLoisGL}


\label{tab:MISEempLois}
\end{table}

\subsection{Application to a real dataset on breast cancer}
\label{sec:RealData}
In this section, we apply the proposed method to study the relapse free survival (RFS) from breast cancer adjusted on high-dimensional covariates in two groups of patients. We consider a Cox model (\ref{eq:CoxIntro34}) to link the RFS to the covariates. We aim at answering the two questions of the introduction concerning the biomarkers that influence the RFS and the prediction of the RFS for each individual.
%
The dataset is available on the website \url{www.ncbi.nlm.nih.gov/geo/query/acc.cgi?acc=GSE6532}. 

The dataset consists of 414 patients in the cohort GSE6532 collected by \citet{Loi} for the purpose of characterizing Estrogen Receptor (ER)-positive subtypes with gene expression profiles. Estrogen receptors are a group of proteins found inside cells, which is activated by the hormone estrogen. There are different forms of estrogen receptors, referred to as subtypes of estrogen receptors. When they are over expressed, they are referred to as ER-positive. The dataset has been studied from a survival analysis point of view in \citet{TibLUMINAL}. Following them, we apply the two  procedures to the same survival time of interest (the RFS).
Excluding patients with incomplete informations, as it is done by \citet{Loi}, there are 142 patients receiving Tamoxifen and 104 untreated patients. It should be underlined that we should do better to handle the missing data, but in this study we also exclude the patients with missing data.
In addition to clinical informations such as the age or the size of the tumor, 
we have 44 928 gene expression measurements for each of the 246 patients.  
Two different survival times are available in this study: the time of relapse free survival and  the time of distant metastasis free survival.
We are interested in this study in the time of relapse free survival, which subjects to right censoring due to incomplete follow-up. There are $60\%$ of censorship in the group of the untreated patients and $66\%$ in the group of patients receiving Tamoxifen. Our goal is to compare the baseline functions in the two groups of patients: the patients receiving Tamoxifen and the untreated patients. 

We start by a preliminary variable selection among the $44928$ levels of gene expression. This corresponds to a screening step (see \citet{Fan10siscox}). This preliminary variable selection is based on the score statistics of each Cox model considered for each variable separately. We only keep the variables which score statistics are superior to a threshold. The difference from the procedure proposed by \citet{Fan10siscox} is that we fix the number of covariates we want to keep and then we tune a threshold to select this number of covariates.
We define the threshold as the $95^{th}$ percentile of a Chi-squared distribution with 1 degree of freedom, so that $996$ probesets have been selected and with the clinical covariates, we have $p=1000$.

\begin{figure}[htp!]
    \centering
    \begin{subfigure}[b]{0.45\textwidth} 
        \includegraphics[height=6.5cm,width=6cm]{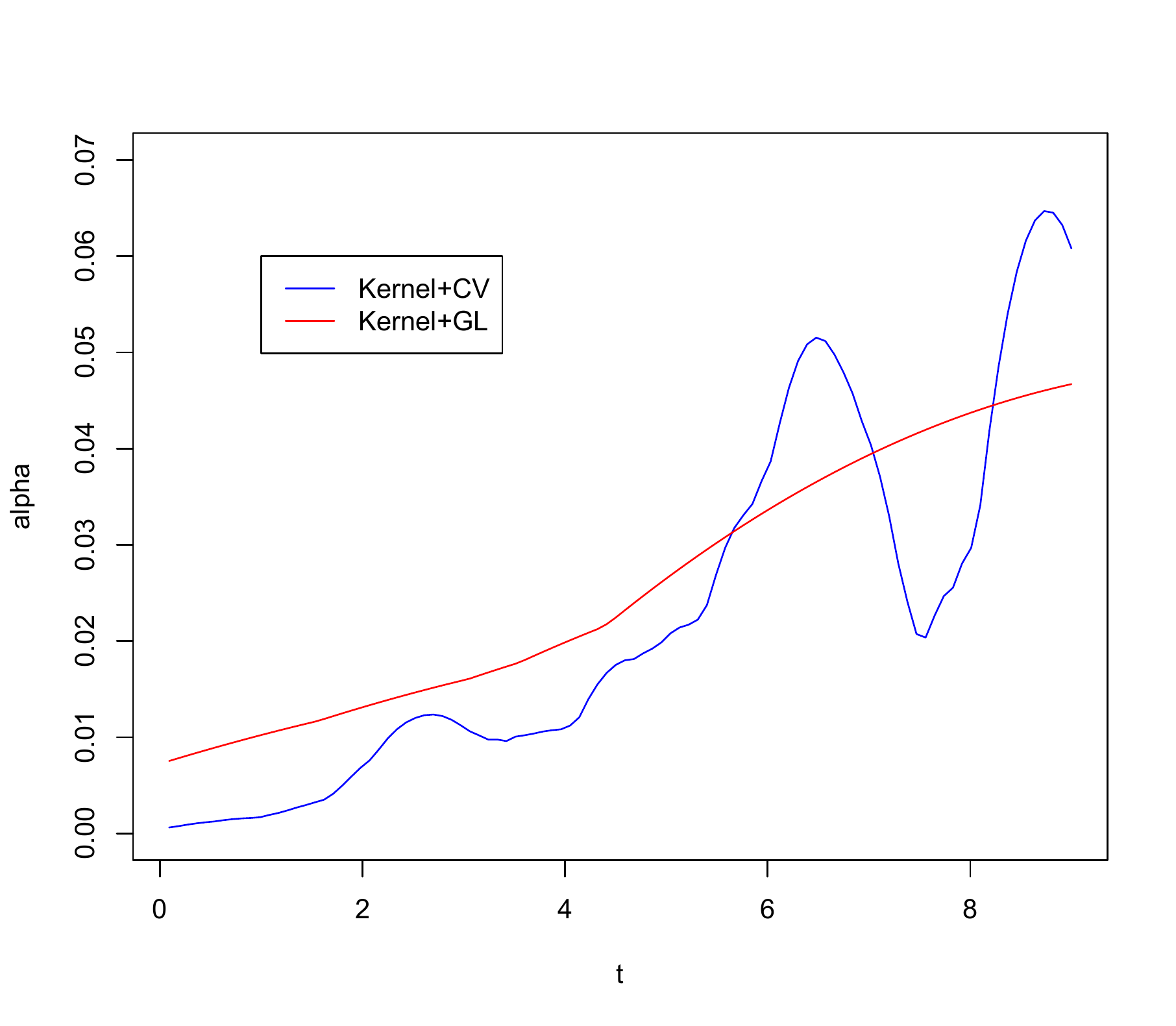} 
       \caption{\hspace{0.2cm}Untreated patients (p=1000).}
        \label{fig:UntreatTrigo1000.}
    \end{subfigure}
    \begin{subfigure}[b]{0.45\textwidth}
        \centering 
        \includegraphics[height=6.5cm,width=6cm]{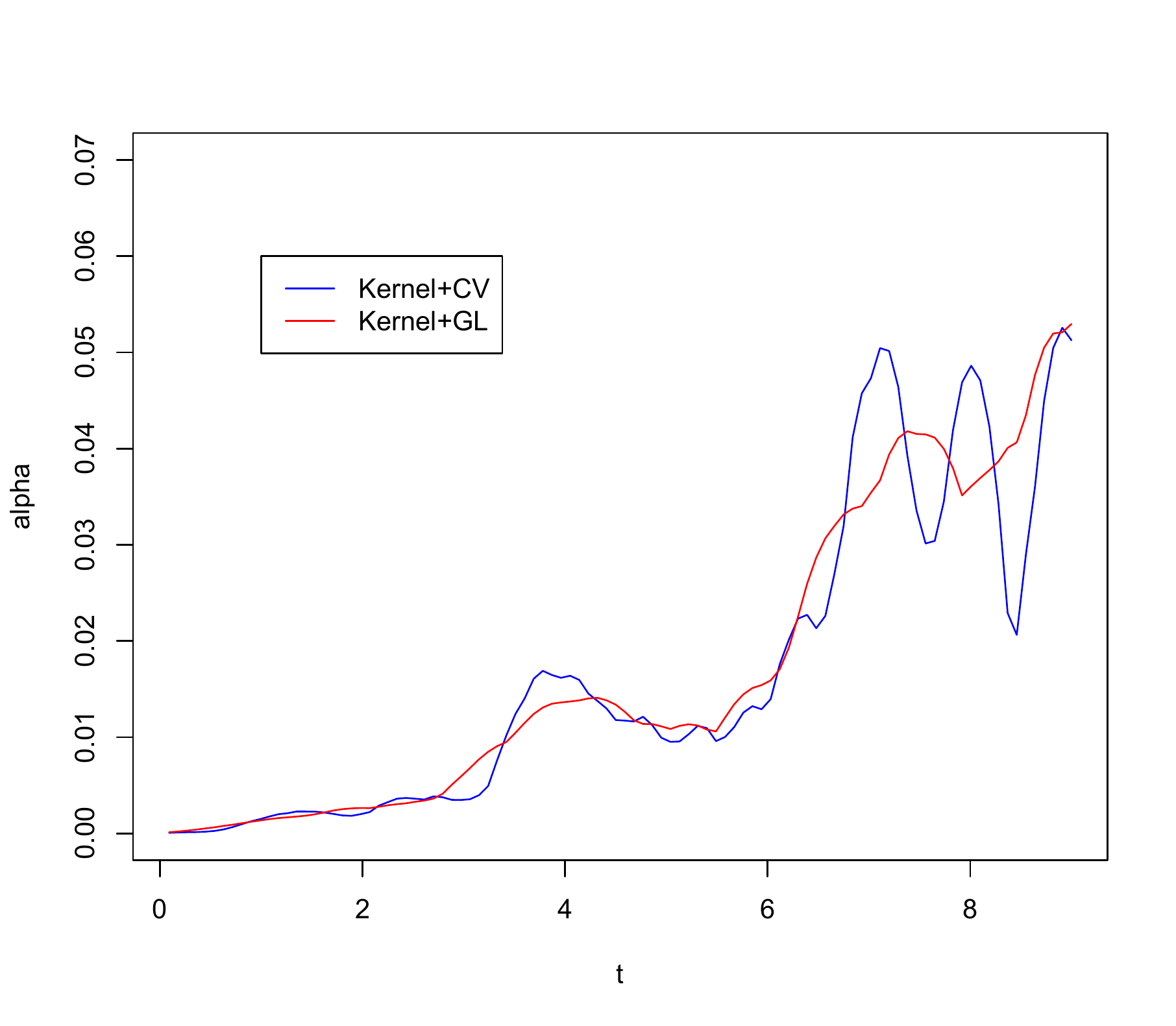}
        \caption{\hspace{0.5cm}Tamoxifen patients (p=1000).}
        \label{fig:TamoxTrigo1000.}
    \end{subfigure}

    \vspace{0.5cm}
    \caption{Kernel estimator with a bandwidth selected by cross-validation (in blue) and kernel estimator with a bandwidth selected with the Goldenshluger and Lepski method (in red). The righthand plot is associated to the group of untreated patients and the lefthand plot correspond to the group of Tamoxifen patients for $p=1000$.}
\label{fig:courbesDRTrigo}
\end{figure}

Figure \ref{fig:courbesDRTrigo} shows the graphs of the kernel estimators of the baseline function with a bandwidth selected by cross-validation and by the Goldenshluger and Lepski method, in the two groups of patients for $p=1000$.

On Figure \ref{fig:courbesDRTrigo}, we observe that the estimator obtained by cross-validation fails to give an interpretable estimate of $\alpha_0$ for the untreated patients. For the estimator obtained from the Goldenshluger and Lepski method, we observe that the risk of relapse to breast cancer has slowed down with the treatment, because the estimated baseline function is close to 0 until $t=2.5$ for the patients treated with tamoxifen whereas it already increases at time $t=1.5$ for the untreated patients.
This leads us to believe that the treatment has a positive influence on the survival time. 

\section{Proofs}
\label{proof}

This section is organized as follows. First, we establish a lemma that allows to control the estimation error of the kernel estimator for a fixed bandwidth $h$, then we prove Theorem \ref{th:hatalphahath-alpha0} from two fundamental lemmas 
that are also proved in this section. We add a supplementary material for all the other used technical lemmas, that are not essential for a first reading.  


\subsection{Intermediate lemma: bound for the kernel estimator of $\alpha_0$ with a fixed bandwidth}
\label{proof:pseudo}

We first establish a non-asymptotic global bound on Mean Integrated Squared Error ($\MISE$) for the estimators $\hat\alpha^{\bs{\hat\beta}}_h$, with $h$ fixed.


\begin{lemma}
\label{th:hatalpha-alpha0}
Under Assumptions \ref{ass:betaball}, \ref{ass:Z},  \ref{ass:baseline3}.\ref{ass:cS3}-\ref{ass:alpha0inf3} and \ref{ass:kernel3}.\ref{ass:kerinf}-\ref{ass:ConvKernel}, for a fixed $h>0$, $n$ large enough and $k\geq 12$
\begin{equation}
\label{eq:hatalpha-alpha0}
\E[||\hat\alpha^{\bs{\hat\beta}}_h-\alpha_0||^2_{2}]\leq 2||\alpha_h-\alpha_0||^2_{2}+\dfrac{C_1}{nh}+C_2(s)\dfrac{\log(pn^k)}{n}
\end{equation}
where $C_1$ is a constant depending on $\tau$, $\alphainf$, $c_S$, $\E[\e^{\bs{\beta_0^TZ_1}}]$, $\E[\e^{2\bs{\beta_0^TZ_1}}]$, $\tau$ and $||K||_{\Lint}$ and  $C_2(s)$ is a constant depending on $B$, $|\bs{\beta_0}|_1$, $R$, $\alphainf$, $c_S$, $\tau$, $||K||_{\Lint}$  and on the sparsity index $s$ of $\bs{\beta_0}$.
\end{lemma}


To prove this lemma and link the kernel estimator to the true baseline function $\alpha_0$, the trick is to introduce a pseudo-estimator, which does not depend on $\bs{\hat\beta}$. Consider for $h>0$ the pseudo-estimator 

\begin{equation}
\label{pseudo}
\bar\alpha_h(t)=\dfrac{1}{nh}\sumin\inttau K\Big(\dfrac{t-u}{h}\Big)\dfrac{1}{S(u,\bs{\beta_0})}\diff N_i(u),
\end{equation}
which corresponds to the kernel estimator of $\alpha_0$ when $S(u,\bs{\beta_0})=\mathbb{E}[\e^{\bs{\beta_0^TZ_i}}Y_i(u)]$ is known. 
To justify the choice of the pseudo-estimator, let us calculate its expectation:
\begin{align*}
\mathbb{E}[\bar\alpha_h(t)]&=\dfrac{1}{nh}\sumin\int_0^{\tau} K\Big(\dfrac{t-u}{h}\Big)\dfrac{1}{S(u,\bs{\beta_0})}\alpha_0(u)\mathbb{E}[\expbeta Y_i(u)]\diff u\\
&=\dfrac{1}{h}\int_0^\tau K\Big(\dfrac{t-u}{h}\Big)\alpha_0(u)\diff u\\
&=K_h*\alpha_0(t),
\end{align*}
which is a unit approximation of $\alpha_0$, so that $K_h*\alpha_0\underset{h\rightarrow 0}{\longrightarrow}\alpha_0$ under mild conditions (see Bochner Lemma and Assumption \ref{ass:kernel3}.\ref{ass:ConvKernel}).\\
In the following, we define for all $t\in[0,\tau]$ 
\begin{align}
\label{alphah}
\alpha_h(t):=\E[\bar\alpha_h(t)]=K_h*\alpha_0(t).
\end{align}

The proof is based on the following decomposition for $h>0$
\begin{align}
\label{eq:decomp}
\E[||\hat\alpha^{\bs{\hat\beta}}_h-\alpha_0||^2_{2}]\leq 2\E[||\hat\alpha^{\bs{\hat\beta}}_h-\bar\alpha_h||^2_{2}]+ 2\E[||\bar\alpha_h-\alpha_0||^2_{2}].
\end{align}
Since the pseudo-estimator (\ref{pseudo}) does not depend on the estimator $\bs{\hat\beta}$, the error $\E[||\bar\alpha_h-\alpha_0||^2_{2}]$ is easier to bound than directly the error $\E[||\hat\alpha^{\bs{\hat\beta}}_h-\alpha_0||^2_{2}]$.  
The study of the error of $\hat\alpha^{\bs{\hat\beta}}_h-\alpha_0$ is then divided into two parts: the error of $\bar\alpha_h-\alpha_0$ and the one of $\hat\alpha^{\bs{\hat\beta}}_h-\bar\alpha_h$.
\newline

The following lemma provides the classical bias/variance inequality for the pseudo-estimator (\ref{pseudo}).
\begin{lemma}
\label{prop:tildealpha-alpha0}
Under Assumptions \ref{ass:baseline3}.\ref{ass:cS3}-\ref{ass:alpha0inf3}, \ref{ass:kernel3}.\ref{ass:kerinf}-\ref{ass:ConvKernel}, for $h>0$ fixed
\begin{align}
\label{th:tildealpha-alpha0}
\mathbb{E}[||\bar\alpha_h-\alpha_0||^2_{2}]\leq ||\alpha_h-\alpha_0||^2_{2}+\dfrac{2\alphainf\tau}{c_S^2}\Big(\E[\e^{\bs{\beta_0^TZ_1}}]+\alphainf\E[\e^{2\bs{\beta_0^TZ_1}}]\tau\Big)\dfrac{||K||_{\Lint}^2}{nh}.
\end{align}
\end{lemma}
The next lemma controls the quadratic error between $\hat\alpha^{\bs{\hat\beta}}_h$ and $\bar\alpha_h$. The term to be controlled in this difference is in fact the difference between the regression parameter $\bs{\beta_0}$ and its Lasso estimator $\bs{\hat\beta}$. The $\ell_1$-norm of this difference is bounded from Proposition \ref{IObeta3} by a term of order $\log(np)/n$. This explains the obtained bound in the following lemma.
\begin{lemma}
\label{lem:hatalpha-tildealpha}
Under Assumptions \ref{ass:baseline3}.\ref{ass:cS3}-\ref{ass:alpha0inf3}, \ref{ass:kernel3}.\ref{ass:kerinf}-\ref{ass:ConvKernel}, \ref{ass:betaball} and \ref{ass:Z}, for a fixed $h>0$,
\[
\E[||\hat\alpha^{\bs{\hat\beta}}_h-\bar\alpha_h||^2_{2}]\leq c(s)\dfrac{\log(n^kp)}{n},
\]
where $c(s)$ is a constant depending on $B$, $|\bs{\beta_0}|_1$, $R$, $\alphainf$, $c_S$, $\tau$, $||K||_{\Lint}$ and $s$ the sparsity index of $\bs{\beta_0}$.
\end{lemma}

From Equation (\ref{eq:decomp}), gathering Lemmas \ref{prop:tildealpha-alpha0} and \ref{lem:hatalpha-tildealpha} provide directly Lemma \ref{th:hatalpha-alpha0}.

Lemmas \ref{prop:tildealpha-alpha0} and \ref{lem:hatalpha-tildealpha} are proved in the supplementary material.

\subsection{Proof of the oracle inequality in Theorem \ref{th:hatalphahath-alpha0}}

%


For all $h\in\mathcal{H}_n$, $A^{\bs{\hat\beta}}(h)$ is defined by (\ref{eq:Ah}) and we can apply this definition for $h=\hat h^{\bs{\hat\beta}}$. We deduce from this, using Definition (\ref{eq:hath3}) of $\hat h^{\bs{\hat\beta}}$, that for all $h\in\mathcal{H}_n$
\begin{align*}
||\hat\alpha^{\bs{\hat\beta}}_{\hat h^{\bs{\hat\beta}}}-\alpha_0||^2_{2}&\leq 3||\hat\alpha^{\bs{\hat\beta}}_{\hat h^{\bs{\hat\beta}}}-\hat\alpha^{\bs{\hat\beta}}_{h,\hat h^{\bs{\hat\beta}}}||^2_{2}+3||\hat\alpha^{\bs{\hat\beta}}_{h,\hat h^{\bs{\hat\beta}}}-\hat\alpha^{\bs{\hat\beta}}_h||^2_{2}+3||\hat\alpha^{\bs{\hat\beta}}_h-\alpha_0||^2_{2}\\
&\leq 3(A^{\bs{\hat\beta}}(h)+V(\hat h^{\bs{\hat\beta}}))+3(A^{\bs{\hat\beta}}(\hat h^{\bs{\hat\beta}})+V(h))+3||\hat\alpha^{\bs{\hat\beta}}_h-\alpha_0||^2_{2}\\
&\leq 6(A^{\bs{\hat\beta}}(h)+V(h))+3||\hat\alpha^{\bs{\hat\beta}}_h-\alpha_0||^2_{2}.
\end{align*}
We obtain for $h\in\mathcal{H}_n$
\begin{equation}
\label{eq:AhVh}
\E[||\hat\alpha^{\bs{\hat\beta}}_{\hat h^{\bs{\hat\beta}}}-\alpha_0||^2_{2}]\leq 6\E[A^{\bs{\hat\beta}}(h)]+6V(h)+3\E[||\hat\alpha^{\bs{\hat\beta}}_h-\alpha_0||^2_{2}].
\end{equation}
Lemma \ref{th:hatalpha-alpha0} gives a bound of $\E[||\hat\alpha^{\bs{\hat\beta}}_h-\alpha_0||^2_{2}]$, which reveals the bias term, the variance term of order $1/nh$ and a remaining term of order $\log(np)/n$, and $V(h)$ is of the expected order $1/nh$. $\E[A^{\bs{\hat\beta}}(h)]$ is bounded in the following proposition.

\begin{proposition}
\label{lem:Ah}
Let $h\in\mathcal{H}_n$ be fixed. Under the assumptions of Theorem \ref{th:hatalphahath-alpha0}, there exist constants $C_1$, $C_2(s)$, $C_3(s)$ such that,
\begin{align}
\label{eq:lemAh}
\E[A^{\bs{\hat\beta}}(h)]\leq C_1 ||\alpha_h-\alpha_0||^2_{2}+C_2(s)\dfrac{\log^{a}(n)\log(n^kp)}{n}+C_3(s)\dfrac{\log(n^kp)}{n},
\end{align}
where the constant $C_1$ only depends on $||K||_1$.
\end{proposition}
Applying Inequalities (\ref{eq:hatalpha-alpha0}) and (\ref{eq:lemAh}) in Equation (\ref{eq:AhVh}) implies Inequality (\ref{eq:adaptiveOI}) by taking the infimum over $h\in\mathcal{H}_n$. This ends the proof of Theorem \ref{th:hatalphahath-alpha0}.
\qed

\subsection{Proof of Proposition \ref{lem:Ah}}

%
We introduce several additional notations $\bar\alpha_{h,h'}=K_{h'}*\bar\alpha_h$, $\alpha_h(t)=\E[\bar\alpha_h(t)]$, $\alpha_{h,h'}(t)=\E[\bar\alpha_{h,h'}(t)]$, and write
\begin{align*}
A^{\bs{\hat\beta}}(h)&=\underset{h'\in\mathcal{H}_n}{\sup}\Big\{||\hat\alpha^{\bs{\hat\beta}}_{h'}-\hat\alpha^{\bs{\hat\beta}}_{h,h'}||^2_{2}-V(h')\Big\}_+\\
&\leq 5\underset{h'\in\mathcal{H}_n}{\sup}\Big\{||\bar\alpha_{h'}-\alpha_{h'}||^2_{2}-V(h')/10\Big\}_+\hspace{-0.2cm}+5\underset{h'\in\mathcal{H}_n}{\sup}\Big\{||\bar\alpha_{h,h'}-\alpha_{h,h'}||^2_{2}-V(h')/10\Big\}_+\\
&\hspace{0.4cm}+5\underset{h'\in\mathcal{H}_n}{\sup}||\hat\alpha^{\bs{\hat\beta}}_{h'}-\bar\alpha_{h'}||^2_{2}+5\underset{h'\in\mathcal{H}_n}{\sup}||\hat\alpha^{\bs{\hat\beta}}_{h,h'}-\bar\alpha_{h,h'}||^2_{2}+5\underset{h'\in\mathcal{H}_n}{\sup}||\alpha_{h'}-\alpha_{h,h'}||^2_{2}\\
&:=5(T_1+T_2+T_3+T_4+T_5)
\end{align*}

$\bullet$ {Study of $\E[T_1]$} : 
Recall that for all $h\in\mathcal{H}_n$
\begin{equation}
\label{eq:supPS}
||\bar\alpha_{h}-\alpha_{h}||^2_{2}=\underset{f\in \mathbb{L}^2([0,\tau]), ||f||_{2}=1}{\sup}\langle \bar\alpha_{h}-\alpha_{h}, f \rangle^2_{2}.
\end{equation}
We introduce the centered empirical process $\nu_{n,h}(f)=\langle \bar\alpha_{h}-\alpha_{h}, f \rangle_{2}$, which is equal to
\begin{align*}
\dfrac{1}{n}\sumin\int_{0}^{\tau}f(t)\left(\inttau \dfrac{K_{h}(t-u)}{\s}\left(\diff N_i(u)-\alpha_0(u)\s\diff u\right)\right)\diff t.
\end{align*}
As $f\longmapsto\nu_{n,h}(f)$ is continuous, the supremum in (\ref{eq:supPS}) can be taken over a countable dense subset of $\{f\in \mathbb{L}^2([0,\tau]),||f||_{2}=1\}$, which we denote by $\mathcal{B}_\tau$. Therefore, we can write
\begin{align}
\nonumber
\E[T_1]&\leq \E\Big[\Big\{\underset{h'\in\mathcal{H}_n}{\sup}||\bar\alpha_{h'}-\alpha_{h'}||^2_{2}-V(h')/10\Big\}_+\Big]\\
\nonumber
&\leq\sum_{h'\in\mathcal{H}_n}\E\Big[\Big\{||\bar\alpha_{h'}-\alpha_{h'}||^2_{2}-V(h')/10\Big\}_+\Big]\\
\label{eq:ET1}
&\leq \sum_{h'\in\mathcal{H}_n}\E\left[\left\{\underset{f\in\mathcal{B}_\tau}{\sup}\nu^2_{n}(f)- V(h')/10\right\}_+\right].
\end{align}
%
Let introduce a key lemma, which allows to bound (\ref{eq:ET1}).  
\begin{lemma}
\label{lem:key}
Let us introduced the centered process $\nu_{n,h}(f)=\langle \bar\alpha_{h}-\alpha_{h}, f \rangle_2$, for any $h\in\mathcal{H}_n$ and $f\in\mathbb{L}^2([0,\tau])$ and $\mathcal{B}_\tau=\{f\in \mathbb{L}^2([0,\tau]),||f||_2=1\}$.
Under the assumptions of Theorem \ref{th:hatalphahath-alpha0}, with $V(h')$ defined by (\ref{eq:Vh}) for any $h'\in\mathcal{H}_n$, there exists two constants $c_6$ and $c_7$ depending on the bound $\kappa_b$ of the B\"urkholder Inequality, $\tau$, $||\alpha_0||_{\infty,\tau}$, the bound $c_S$ of $S(t,\bs{\beta_0})$, $\E[\e^{\bs{\beta_0^TZ_1}}]$, $\E[\e^{2\bs{\beta_0^TZ_1}}]$, $||K||_{\mathbb{L}^1(\mathbb{R})}$ and $||K||_{\Lint}$, such that
\[
\sum_{h\in\mathcal{H}_n}\E\Bigg[\Bigg\{\underset{f\in\mathcal{B}_\tau(h)}{\sup}\nu^2_{n,h}(f)- V(h)/10\Bigg\}_+\Bigg]\leq \dfrac{c_6}{n}+c_7\dfrac{\log^a(n)}{n}.
\]
\end{lemma}
So, from Lemma \ref{lem:key}, there exists two constants $c_6>0$ and $c_7>0$ such that
\begin{equation}
\label{eq:T1}
\E[T_1]\leq \dfrac{c_6}{n}+c_7\dfrac{\log^a(n)}{n},
\end{equation}
where $c_6$ and $c_7$ depend on $\tau$, $\alphainf$, $c_S$, $\E[\e^{\bs{\beta_0^TZ_1}}]$, $\E[\e^{2\bs{\beta_0^TZ_1}}]$, $||K||_{\mathbb{L}^1(\mathbb{R})}$ and $||K||_{\Lint}$.
\newline

$\bullet$ {Study of $\E[T_2]$} : We study $T_2$ similarly as  $T_1$ since
\[
\E[T_2]\leq \sum_{h'\in\mathcal{H}_n}\E\left[\left\{||\bar\alpha_{h,h'}-\alpha_{h,h'}||^2_{2,h'}- V(h')/10\right\}_+\right].
\]
From Lemma \ref{lem:key} (see the remark at the end of the proof of Lemma \ref{lem:key}), there exists two constants $c_8>0$ and $c_9>0$ such that 
\begin{equation}
\label{eq:T2}
\E[T_2]\leq \dfrac{c_8}{n}+c_9\dfrac{\log^a(n)}{n},
\end{equation}
where $c_8$ and $c_9$ depend on $\tau$, $\alphainf$, $c_S$, $\E[\e^{\bs{\beta_0^TZ_1}}]$, $\E[\e^{2\bs{\beta_0^TZ_1}}]$, $||K||_{\mathbb{L}^1(\mathbb{R})}$ and $||K||_{\Lint}$.\newline

$\bullet$ {Study of $\E[T_3]$} :
First, write for all $h\in\mathcal{H}_n$, that
\begin{align*}
\hat\alpha^{\bs{\hat\beta}}_h(t)-\bar\alpha_h(t)=\dfrac{1}{nh}\sumin\inttau K\Bigg(\dfrac{t-u}{h}\Bigg)\dfrac{S(u,\bs{\beta_0})\mathds{1}_{\{\bar Y(u)>0\}}-S_n(u,\bs{\hat\beta})}{S_n(u,\bs{\hat\beta})S(u,\bs{\beta_0})}\diff N_i(u)
\end{align*}
For all $u\in[0,\tau]$, we have $S(u,\bs{\beta_0})\mathds{1}_{\{\bar Y(u)>0\}}-S_n(u,\bs{\hat\beta})=(S(u,\bs{\beta_0})-S_n(u,\bs{\hat\beta}))\mathds{1}_{\{\bar Y(u)>0\}}$. Indeed, for all $u\in[0,\tau]$, if $\mathds{1}_{\{\bar Y(u)>0\}}=0$, then for all $i\in\{1,...,n\}$, $Y_i(u)=0$ and $S_n(u,\bs{\hat\beta})=0$. So, we can rewrite for all $h\in\mathcal{H}_n$ that
\begin{align}
\label{eq:diffPseudoEst}
\hat\alpha^{\bs{\hat\beta}}_h(t)-\bar\alpha_h(t)=\dfrac{1}{nh}\sumin\inttau K\Bigg(\dfrac{t-u}{h}\Bigg)\dfrac{S(u,\bs{\beta_0})-S_n(u,\bs{\hat\beta})}{S_n(u,\bs{\hat\beta})S(u,\bs{\beta_0})}\mathds{1}_{\{\bar Y(u)>0\}}\diff N_i(u).
\end{align}
Consider the following sets:
\begin{align}
\label{OmegaHk}
\Omega_{H,k}&=\Big\{\omega : \forall u \in [0,\tau], |S_n(u,\bs{\hat\beta})-S(u,\bs{\beta_0})|\leq 2C(s)B\e^{BR}\e^{2B|\bs{\beta_0}|_1}\sqrt{\dfrac{\log(pn^k)}{n}}\Big\},\\
\label{OmegaSn}
\Omega_{S_n}&=\Big\{\omega : \forall u \in [0,\tau], S_n(u,\bs{\hat\beta})-S(u,\bs{\beta_0})\geq -\dfrac{c_S}{2}\Big\},
\end{align}
\begin{equation}
\label{Omega}
\Omega_k=\Omega_{H,k}\cap\Omega_{S_n}.
\end{equation}
We decompose $T_3$ on $\Omega_k$ and on its complement. 
On $\Omega^c_k$, let introduce the following lemma:
\begin{lemma} 
\label{lem:surOmegac}
Under Assumptions \ref{ass:baseline3}.\ref{ass:cS3}-\ref{ass:alpha0inf3}, \ref{ass:kernel3}.\ref{ass:kerinf}-\ref{ass:ConvKernel}, \ref{ass:betaball} and \ref{ass:Z}, for all $k\in\mathbb{N}$, we have
\[
\E[||\hat\alpha^{\bs{\hat\beta}}_h-\bar\alpha_h||^2_{2}\mathds{1}(\Omega^c_k)]\leq c_3 n^{4-k/2},
\]
where $c_3$ is a constant depending on $B$, $|\bs{\beta_0}|_1$, $R$, $\alphainf$, $c_S$, $\tau$, $||K||_{\infty}$.
Choosing $k\geq 10$ yields $\E[||\hat\alpha^{\bs{\hat\beta}}_h-\bar\alpha_h||^2_{2}]\leq c_3/n$.
\end{lemma}
From Lemma \ref{lem:surOmegac}, 
\begin{align*}
\E\Bigg[\underset{h'\in\mathcal{H}_n}{\sup}||\hat\alpha^{\bs{\hat\beta}}_{h'}-\bar\alpha_{h'}||^2_{2}\mathds{1}(\Omega^c_k)\Bigg]&\leq \sum_{h'\in\mathcal{H}_n}\E[||\hat\alpha^{\bs{\hat\beta}}_{h'}-\bar\alpha_{h'}||^2_{2}\mathds{1}(\Omega^c_k)]\\
&\leq \sum_{h'\in\mathcal{H}_n} c_3 n^{4-k/2},
\end{align*}
which is of order $1/n$ as long as $k\geq 12$.
On the other hand, from (\ref{eq:diffPseudoEst}) on $\Omega_k$, we have
\begin{align*}
\E\Bigg[&\underset{h'\in\mathcal{H}_n}{\sup}\inttau(\hat\alpha^{\bs{\hat\beta}}_{h'}-\bar\alpha_{h'})^2(t)\mathds{1}(\Omega_k)\diff t\Bigg]\\
&\leq \dfrac{16C(s)^2B^2\e^{2BR}\e^{4B|\bs{\beta_0}|_1}}{c_S^2}\dfrac{\log(pn^k)}{n}\E\left[\underset{h'\in\mathcal{H}_n}{\sup}\inttau\Bigg(\inttau\dfrac{|K_{h'}(t-u)|}{\s}\Bigg(\dfrac{1}{n}\sumin\diff N_i(u)\Bigg)\Bigg)^2\right].
\end{align*}
Then, we decompose $N_i=(N_i-\Lambda_i)+\Lambda_i$ to obtain
\begin{align}
\nonumber
\E&\left[\underset{h'\in\mathcal{H}_n}{\sup}\inttau\Bigg\{\inttau\dfrac{|K_{h'}(t-u)|}{\s}\Big(\dfrac{1}{n}\sumin\diff N_i(u)\Big)\Bigg\}^2\diff t\right]\\
\label{eq:Term1}
\leq & \hspace{0.1cm}2\E\Bigg[\underset{h'\in\mathcal{H}_n}{\sup}\inttau\Bigg\{\inttau\dfrac{|K_{h'}(t-u)|}{\s} \Big(\dfrac{1}{n}\sumin\diff N_i(u)-\alpha_0(u)\s\diff u\Big)\Bigg\}^2\diff t\Bigg]\\
\label{eq:Term2}
&+2\underset{h'\in\mathcal{H}_n}{\sup}\inttau\Bigg\{\inttau|K_{h'}(t-u)|\alpha_0(u)\diff u\Bigg\}^2\diff t.
\end{align} 
The term (\ref{eq:Term2}) is bounded by $2\tau\alphainf^2||K||_{\mathbb{L}^1(\mathbb{R})}^2$. Let us bound the term (\ref{eq:Term1}), 
\begin{align*}
&\E\Bigg[\underset{h'\in\mathcal{H}_n}{\sup}\inttau\Bigg\{\inttau\dfrac{|K_{h'}(t-u)|}{\s} \Big(\dfrac{1}{n}\sumin\diff N_i(u)-\alpha_0(u)\s\diff u\Big)\Bigg\}^2\diff t\Bigg]\\
&\leq \sum_{h'\in\mathcal{H}_n}\inttau\var\Bigg[\inttau\dfrac{|K_{h'}(t-u)|}{\s}\dfrac{1}{n}\sumin\diff N_i(u)\Bigg]
\end{align*}
It remains to bound the variance term.
\begin{align*}
\var\Bigg[\dfrac{1}{n}\sumin\inttau\dfrac{|K_{h}(t-u)|}{\s}\diff N_i(u)\Bigg]&\leq \dfrac{1}{n}\mathbb{E}\Bigg[\Bigg(\inttau \dfrac{|K_{h}(t-u)|}{\s}\diff N_1(u)\Bigg)^2\Bigg].
\end{align*}
We apply the Doob-Meyer decomposition to get
\begin{align}
\label{eq:varterm1}
\var\Bigg[\dfrac{1}{n}\sumin\inttau\dfrac{|K_{h}(t-u)|}{\s}\diff N_i(u)\Bigg]&\leq\dfrac{2}{n}\E\Bigg[\Bigg(\inttau \dfrac{K_{h}(t-u)}{S(u,\bs{\beta_0})}\diff M_1(u)\Bigg)^2\Bigg]\\
\label{eq:varterm2}
&+\dfrac{2}{n}\E\Bigg[\Bigg(\inttau \dfrac{K_{h}(t-u)}{S(u,\bs{\beta_0})}\alpha_0(u)\e^{\bs{\beta_0^TZ_1}}Y_1(u)\diff u\Bigg)^2\Bigg].
\end{align}
The term (\ref{eq:varterm1}) is bounded by 
\begin{align}
\label{eq:varterm11}
\dfrac{2}{n}\E\Bigg[\inttau\dfrac{K^2_{h'}(t-u)}{(S(u,\bs{\beta_0}))^2}\alpha_0(u)\e^{\bs{\beta_0^TZ_1}}Y_1(u)\diff u\Bigg]\leq \dfrac{2}{n}\dfrac{\alphainf\E[\e^{\bs{\beta_0^TZ_1}}]}{c_S^2}\dfrac{||K||_{\Lint}^2}{h'},
\end{align}
and from the Cauchy-Schwarz inequality, the term (\ref{eq:varterm2}) is bounded by
\begin{align}
\label{eq:varterm22}
\dfrac{2}{n}\dfrac{\alphainf^2\E[\e^{2\bs{\beta_0^TZ_1}}]\tau}{c_S^2}\dfrac{||K||_{\Lint}^2}{h'}.
\end{align}
From (\ref{eq:varterm11}) and (\ref{eq:varterm22}), (\ref{eq:Term1}) is bounded by
\begin{align}
\label{eq:varterm3}
\dfrac{4}{n}\dfrac{\alphainf\tau}{c_S^2}\Big[\E[\e^{\bs{\beta_0^TZ_1}}]+\alphainf\E[\e^{2\bs{\beta_0^T}}]\Big]||K||_{\Lint}^2\sum_{h'\in\mathcal{H}_n}\dfrac{1}{nh'}.
\end{align}
From Condition \ref{ass:kernel3}.\ref{eq:sum1/nh} and bounds (\ref{eq:varterm2}) and (\ref{eq:varterm3}), we deduce that 
\begin{align*}
\E\Bigg[\underset{h'\in\mathcal{H}_n}{\sup}\inttau&(\hat\alpha^{\bs{\hat\beta}}_{h'}-\bar\alpha_{h'})^2(t)\mathds{1}(\Omega_k)\diff t\Bigg]\\
&\leq C(s,c_S,B,R,|\bs{\beta_0}|_1,\alphainf,\tau,||K||_{\Lint},\E[\e^{\bs{\beta_0^TZ_1}}],\E[\e^{2\bs{\beta_0^TZ_1}}])\dfrac{\log^a(n)\log(pn^k)}{n}.
\end{align*}
Finally, there exists a constant $c_5>0$ such that
\begin{equation}
\label{eq:T3}
\E[T_3]\leq c_5\dfrac{\log^{a}(n)\log(n^kp)}{n},
\end{equation}
where $c_5$ depends on $s$, $c_S$, $B$, $R$, $\tau$, $\alphainf$, $|\bs{\beta_0}|_1$, $||K||_{\Lint}$, $\E[\e^{\bs{\beta_0^TZ_1}}]$ and $\E[\e^{2\bs{\beta_0^TZ_1}}]$.
\newline
 
$\bullet$ {Study of $\E[T_4]$} :
Since 
\[
\hat\alpha^{\bs{\hat\beta}}_{h,h'}-\bar\alpha_{h,h'}=K_{h'}*(\hat\alpha^{\bs{\hat\beta}}_h-\bar\alpha_h),
\]
we have from Young Inequality (Lemma 2.2 in the supplementary material) with $p=1,q=2,r=2$,
\begin{equation}
\label{eq:T4}
\E[T_4]\leq ||K||^2_{\mathbb{L}^1(\mathbb{R})}\E[||\hat\alpha^{\bs{\hat\beta}}_h-\bar\alpha_h||^2_{2}]\leq C(s)||K||^2_{\mathbb{L}^1(\mathbb{R})}\dfrac{\log (n^kp)}{n},
\end{equation}
where the last inequality is obtained from Lemma \ref{lem:hatalpha-tildealpha}.
\newline

$\bullet$ {Study of $\E[T_5]$} :
From Young Inequality (Lemma 2.2 in the supplementary material) with $p=1,q=2,r=2$, we obtain that
\[
||\alpha_{h'}-\alpha_{h,h'}||^2_{2}=||K_{h'}*(\alpha_0-K_h*\alpha_0)||^2_{2}\leq ||K||^2_{\mathbb{L}^1(\mathbb{R})}||\alpha_0-K_h*\alpha_0||^2_{2}
\]
Therefore, since $\alpha_h=K_h*\alpha_0$,
\begin{equation}
\label{eq:T5}
\E[T_5]\leq ||K||^2_{\mathbb{L}^1(\mathbb{R})}||\alpha_0-\alpha_h||^2_{2},
\end{equation}
which corresponds to a bias term.
\newline

Finally, gathering the bounds of the five terms (\ref{eq:T1}), (\ref{eq:T2}), (\ref{eq:T3}), (\ref{eq:T4}) and (\ref{eq:T5}), gives the result of Proposition \ref{lem:Ah}.\qed

\subsection{Proof of Lemma \ref{lem:key}}
\label{eq:Tala}
We have to control the supremum of $\nu_{n,h}(f)$ defined by (\ref{eq:nunh}) over the ball $\mathcal{B}_\tau=\{g\in\mathbb{L}^2([0,\tau]), ||g||_{2}=1\}$. 
For all $h\in\mathcal{H}_n$ and $f\in\mathcal{B}_\tau$, we have 
\begin{align}
\label{eq:nunh}
\nu_{n,h}(f)=\dfrac{1}{n}\sumin\int_{0}^{\tau}f(t)\left(\displaystyle\int_0^\tau \dfrac{K_{h}(t-u)}{\s}\left(\diff N_i(u)-\alpha_0(u)\s\diff u\right)\right)\diff t.
\end{align}
Usually, to control such a process, we apply the Talagrand Inequality given in Theorem \ref{th:Talagrand}.
However, since $\nu_{n,h}(f)$ is not bounded, we can not directly apply the Talagrand Inequality: we have to introduce a truncation (see \citet{chagny2013} for a close approach).
 Let us define for a constant $c$,
\[
\kappa_n=c\dfrac{\sqrt{n}}{\log n}, 
\]
and we decompose $\nu_{n,h}$ as
\begin{align*}
\nu_{n,h}(f)=\nu_{n,h}^{(1)}(f)+\nu_{n,h}^{(2)}(f),
\end{align*}
where 
\begin{align*}
\nu_{n,h}^{(1)}(f)&=\dfrac{1}{n}\sumin\int_{0}^{\tau} f(t)\displaystyle\int_0^\tau \dfrac{K_{h}(t-u)}{\s}\mathds{1}_{\{N_i(\tau)\leq \kappa_n\}}\diff N_i(u)\diff t\\
&-\dfrac{1}{n} \sumin\int_{0}^{\tau}  f(t)\displaystyle\int_0^\tau\E\Bigg[\dfrac{K_{h}(t-u)}{\s}\mathds{1}_{\{N_i(\tau)\leq \kappa_n\}}\diff N_i(u)\Bigg]\diff t,
\end{align*}
and
\begin{align*}
\nu_{n,h}^{(2)}(f)&=\dfrac{1}{n}\sumin\int_{0}^{\tau} f(t)\displaystyle\int_0^\tau \dfrac{K_{h}(t-u)}{\s}\mathds{1}_{\{N_i(\tau)> \kappa_n\}}\diff N_i(u)\diff t\\
&-\dfrac{1}{n} \sumin\int_{h}^{\tau-h}  f(t)\displaystyle\inttau\E\Bigg[\dfrac{K_{h}(t-u)}{\s}\mathds{1}_{\{N_i(\tau)> \kappa_n\}}\diff N_i(u)\Bigg]\diff t.
\end{align*}
%

\begin{itemize}
\item[$\bullet$] Control of $\nu_{n,h}^{(1)}(f)$: \\
We can apply a Talagrand Inequality to $\nu_{n,h}^{(1)}(f)$, which is bounded.
To apply this concentration inequality, we need to determine the bounds $H$, $M$, $W$ and the constant $\varepsilon$ (see Theorem \ref{th:Talagrand} in Appendix \ref{appendix:TechnicalLemma3} for the notations). 

\begin{itemize}
\item Determination of the constant $M$:\\
Using the Cauchy-Schwarz Inequality, we have for 
\begin{align*}
\Bigg|\int_{0}^{\tau} f(t)\displaystyle\int_{0}^{\tau} &\dfrac{K_{h}(t-u)}{\s}\mathds{1}_{\{N_1(\tau)\leq \kappa_n\}}\diff N_1(u)\diff t\Bigg|\\
\leq &||f||_2\Bigg|\displaystyle\int_{0}^{\tau}\Big(\int_{0}^{\tau} K^2_{h}(t-u)\diff t\Big)^{1/2}\dfrac{\mathds{1}_{\{N_1(\tau)\leq \kappa_n\}}}{\s}\diff N_1(u)\Bigg|\\
\leq &\dfrac{||K||_{\Lint}^2}{\sqrt{h}}\dfrac{|N_1(\tau)\mathds{1}_{\{N_1(\tau)\leq \kappa_n\}}|}{c_S}\\
\leq &\dfrac{||K||_{\Lint}^2}{c_S\sqrt{h}}\kappa_n:=M\sim\dfrac{\sqrt{n}}{\log n\sqrt{h}}.
\end{align*}

\item Determination of the constant $H$:\\
Let define
\[
\psi_h(t)=\dfrac{1}{n}\sumin\inttau \dfrac{K_h(t-u)}{S(u,\bs{\beta_0})}\mathds{1}_{\{N_i(\tau)\leq \kappa_n\}}\diff N_i(u) 
\]
We have $\sup_{f\in\mathcal{B}_\tau}(\nu_{n,h}^{(1)}(f))^2=\sup_{f\in\mathcal{B}_\tau}\langle \psi_h-\E[\psi_h],f \rangle_2^2=||\psi_h-\E[\psi_h]||_2^2$. We deduce from the Doob-Meier decomposition that
\begin{align*}
\E\Big[\underset{f\in\mathcal{B}_\tau}\sup(\nu_{n,h}^{(1)}(f))^2\Big]&=\inttau \var[\psi_h(t)]\diff t\\
&\leq \dfrac{1}{n}\inttau\E\Bigg[\Bigg(\inttau\dfrac{K_h(t-u)}{S(u,\bs{\beta_0})}\mathds{1}_{\{N_1(\tau)\leq \kappa_n\}}\diff N_1(u)\Bigg)^2\Bigg]\diff t\\
&\leq \dfrac{2\alphainf\tau}{c_S^2}\Big(\alphainf\E[\e^{2\bs{\beta_0^TZ_1}}]\tau+\E[\e^{\bs{\beta_0^TZ_1}}]\Big)\dfrac{||K||_{\Lint}^2}{nh}:=H^2
\end{align*}

We have $H^2= V(h)/\kappa$. Then, we set $\varepsilon^2=1/2$ and $\kappa=80$ in order to have $2(1+2\varepsilon^2)H^2= V(h)/20=O(1/nh)$.

\item Determination of the constant $W$:\\
Since $f\in\mathcal{B}_\tau$, we have
\begin{align*}
\var\Bigg[\int_{0}^{\tau}&f(t)\inttau\dfrac{K_{h}(t-u)}{\s}\mathds{1}_{\{N_1(\tau)\leq \kappa_n\}}\diff N_1(u)\diff t\Bigg]\\
\leq&\E\Bigg[\Bigg(\inttau f(t)\inttau\dfrac{K_{h}(t-u)}{\s}\mathds{1}_{\{N_1(\tau)\leq \kappa_n\}}\diff N_1(u)\diff t\Bigg)^2\Bigg]\\
\leq &\E\Bigg[\mathds{1}_{\{N_1(\tau)\leq \kappa_n\}}\Bigg(\inttau\dfrac{(K_h*f)(u)}{\s}\diff N_1(u)\Bigg)^2\Bigg].
\end{align*}
So, from the Doob-Meier decomposition and Young Lemma 2.2 in the supplementary material, we have
\begin{align*}
\var\Bigg[\int_{0}^{\tau}&f(t)\inttau\dfrac{K_{h}(t-u)}{\s}\mathds{1}_{\{N_1(\tau)\leq \kappa_n\}}\diff N_1(u)\diff t\Bigg]\\
\leq &\dfrac{2\alphainf}{c_S^2}\Big[\tau\alphainf\E[\e^{2\bs{\beta_0^TZ_1}}]+\E[\e^{\bs{\beta_0^TZ_1}}]\Big]||K_h*f||_2^2\\
\leq &\dfrac{2\alphainf}{c_S^2}\Big[\tau\alphainf\E[\e^{2\bs{\beta_0^TZ_1}}]+\E[\e^{\bs{\beta_0^TZ_1}}]\Big]||K||_{\mathbb{L}^1(\mathbb{R})}^2:=W
\end{align*}
\end{itemize}

Then, from Assumptions \ref{ass:gridHn}.\ref{eq:sum1/nh} and \ref{eq:sumexp1/h},
we deduce that
\begin{align}
\nonumber
{\sum_{h\in\mathcal{H}_n}\E\Bigg[\Bigg\{\underset{f\in\mathcal{B}_\tau}{\sup}(\nu^{(1)}_{n,h}(f))^2- V(h)/20\Bigg\}_+\Bigg]}&{\leq \dfrac{\vartheta_1}{n}\sum_{h\in\mathcal{H}_n}\Bigg(\e^{-\frac{\vartheta_2}{h}}+\dfrac{1}{n\log^2(n)h}\e^{-\vartheta_3\log{n}}\Bigg)}\\
\label{eq:nu1}
&\leq \dfrac{\tilde\vartheta_1}{n}+\tilde\vartheta_2\dfrac{\log^{a-2}n}{n^{\tilde\vartheta_3}}
\end{align}
with 
\[
V(h)= \kappa\dfrac{2\alphainf\tau}{c_S^2}\Big(\alphainf\E[\e^{2\bs{\beta_0^TZ_1}}]\tau+\E[\e^{\bs{\beta_0^TZ_1}}]\Big)\dfrac{||K||_{\Lint}^2}{nh}.
\]

\item[$\bullet$] Control of $\nu_{n,h}^{(2)}(f)$:\\
Now, let us focus on the second unbounded term $\nu_{n,h}^{(2)}(f)$. Let us consider the process $\Psi(t)$ defined as
\[
\dfrac{1}{n}\sumin\Bigg[\inttau\dfrac{K_{h}(t-u)}{\s}\mathds{1}_{\{N_i(\tau)> \kappa_n\}}\diff N_i(u)-\E\Bigg[\inttau\dfrac{K_{h}(t-u)}{\s}\mathds{1}_{\{N_i(\tau)> \kappa_n\}}\diff N_i(u)\Bigg]\Bigg],
\]
so that $\nu_{n,h}^{(2)}(f)=\int_{0}^{\tau} f(t)\Psi(t)\diff t$.
Using Cauchy-Schwarz inequality, we get
\begin{align*}
\E\Bigg[\underset{f\in\mathcal{B}_\tau}{\sup}(\nu_{n,h}^{(2)}(f))^2\Bigg]&\leq \E\Bigg[\int_{0}^{\tau}\Psi^2(t)\diff t\Bigg]\\
&\leq \dfrac{1}{n}\int_{0}^{\tau}\var\Bigg[\inttau\dfrac{K_{h}(t-u)}{\s}\mathds{1}_{\{N_1(\tau)> \kappa_n\}}\diff N_1(u)\Bigg]\diff t\\
&\leq \dfrac{1}{n}\int_{0}^{\tau}\E\Bigg[\Bigg(\inttau\dfrac{K_{h}(t-u)}{\s}\mathds{1}_{\{N_1(\tau)> \kappa_n\}}\diff N_1(u)\Bigg)^2\Bigg]\diff t
\end{align*}
Applying the Cauchy-Schwarz Inequality (see Lemma 2.1 in the supplementary material), we obtain that for all $k>0$, 
\begin{align*}
\E\Bigg[\underset{f\in\mathcal{B}_\tau}{\sup}(\nu_{n,h}^{(2)}(f))^2\Bigg]& \leq \dfrac{1}{n}\inttau\E\Bigg[\mathds{1}_{\{N_1(\tau)>\kappa_n\}} N_1(\tau)\inttau\dfrac{K_h^2(t-u)}{S^2(u,\bs{\beta_0})}\diff N_1(u)\Bigg]\diff t\\
&\leq \dfrac{||K||_{\Lint}^2}{nhc_S^2}\E[N_1^2(\tau)\mathds{1}_{\{N_1(\tau)>\kappa_n\}}]\\
&\leq \dfrac{||K||_{\Lint}^2}{nhc_S^2}\dfrac{\E[N_1^{k+2}(\tau)]}{\kappa_n^k}\\
&\leq \dfrac{||K||_{\Lint}^2}{nhc_S^2}\dfrac{\E[N_1^{k+2}(\tau)]}{n}
\end{align*}
From Assumption \ref{ass:gridHn}.\ref{eq:sum1/nh}, we deduce that for $k$ large enough
\[
\sum_{h\in\mathcal{H}_n}\E\Bigg[\underset{f\in\mathcal{B}_\tau}{\sup}(\nu_{n,h}^{(2)}(f))^2\Bigg]\leq C\dfrac{\log^a(n)\E[N(\tau)^{k+1}]}{n}.
\]
It remains to verify that $\E[N(\tau)^{k+1}]$ is bounded. Using the fact that for all $a\geq 0$, $b\geq 0$
and $p\geq 1$, $(a+b)^k\leq 2^{k-1}(a^k+b^k)$ and from the B\"urkholder Inequality, we can easily show by recurrence that for all $p\in\mathbb{N}^*$, $\E[N(\tau)^k]\leq C_k$. 
Thus, we conclude that for a good choice of $p$,
\begin{align}
\label{eq:nu2}
\sum_{h\in\mathcal{H}_n}\E\Bigg[\underset{f\in\mathcal{B}_\tau}{\sup}(\nu_{n,h}^{(2)}(f))^2\Bigg]\leq \tilde{C}\dfrac{\log^a(n)}{n},
\end{align}
for a constant $\tilde{C}>0$.

\end{itemize}

Combining (\ref{eq:nu1}) and (\ref{eq:nu2}), we finally get
\begin{equation*}
\sum_{h\in\mathcal{H}_n}\E\Bigg[\Bigg\{\underset{f\in\mathcal{B}_\tau}{\sup}\nu^2_{n,h}(f)- V(h)/10\Bigg\}_+\Bigg]\leq \dfrac{c_6}{n}+c_7\dfrac{\log^a(n)}{n},
\end{equation*}
where $c_6$ and $c_7$ depends on $\tau$, $\alphainf$, $c_S$, $\E[\e^{\bs{\beta_0^TZ_1}}]$, $\E[\e^{2\bs{\beta_0^TZ_1}}]$, $||K||_{\mathbb{L}^1(\mathbb{R})}$ and $||K||_{\Lint}$.
\qed

\paragraph{Remark:} A similar lemma can be obtained for the centered process $\langle\bar\alpha_{h,h'}-~\alpha_{h,h'},f\rangle_{2}$, where $\bar\alpha_{h,h'}=K_{h'}*\bar\alpha_h$ and $\alpha_{h,h'}=\E[\bar\alpha_{h,h'}]$ for $h,h'\in\mathcal{H}_n$. Indeed, from Young Lemma 2.2 in the supplementary material, we have
\begin{align*}
\langle \bar\alpha_{h,h'}-\alpha_{h,h'},f\rangle_2&= \inttau f(t)\Big(K_{h'}*\bar\alpha_h(t)-\E[K_{h'}*\bar\alpha_h(t)]\Big)\diff t\\
&\leq ||f||_2||K_{h'}*(\bar\alpha_h-\E[\bar\alpha_h])||_2\\
&\leq ||f||_2||K||_{\mathbb{L}^1(\mathds{R})}||\bar\alpha_h-\E[\bar\alpha_h]||_2^2.
\end{align*}
Just take the same constants $M$, $H^2$ and $W$ than previously and multiply them by $||K||_{\mathbb{L}^1(\mathds{R})}$.

%

\paragraph{}The proofs of Lemmas \ref{prop:tildealpha-alpha0}, \ref{lem:hatalpha-tildealpha} and \ref{lem:surOmegac} are available in the supplementary material.



%

\bibliography{biblio3}
\bibliographystyle{plainnatfrench}
\end{document}